\documentclass[11pt,a4paper]{article}

\pdfoutput=1
\usepackage{jcappub}
\usepackage{rotating}
\usepackage{array}
\usepackage{amsmath}
\usepackage[normalem]{ulem}
\usepackage{slashed}
\usepackage{bm}
\usepackage{booktabs}
\usepackage[pdftex,table,dvipsnames]{xcolor}
\usepackage{units}
\usepackage[all]{hypcap}
\usepackage{float}
\usepackage{subcaption}
\usepackage[Export]{adjustbox}

\makeatletter
\newcommand{\thickbar}{\mathpalette\@thickbar}
\newcommand{\@thickbar}[2]{{#1\mkern1.2mu\vbox{
  \sbox\z@{$#1\mkern-1.2mu#2\mkern-1.2mu$}%
  \sbox\tw@{$#1\overline{#2}$}%
  \dimen@=\dimexpr\ht\tw@-\ht\z@-2.7\p@\relax
  \hrule\@height1.5\p@ 
  \vskip\dimen@
  \box\z@}\mkern1.5mu}
}
\makeatother

\DeclareFontFamily{U}{mathx}{\hyphenchar\font45}
\DeclareFontShape{U}{mathx}{m}{n}{<-> mathx10}{}
\DeclareSymbolFont{mathx}{U}{mathx}{m}{n}
\DeclareMathAccent{\widebar}{0}{mathx}{"73}

\newcommand{\DRN}{$\widebar{\mbox{\textsc{D}}}$\textsc{arkRayNet}}
\newcommand{\sensitivity}{\mathrm{sens}}
\newcommand{\exper}{\mathrm{exp}}

\usepackage{multirow}

\usepackage[pdftex]{pict2e}

\newcommand{\diff}{\mathrm{d}}

\newcommand{\pbar}{\bar{p}}
\newcommand{\dbar}{\bar{d}}
\newcommand{\tprop}{\boldsymbol{\theta}_\mathrm{prop}}
\newcommand{\tpropi}{\boldsymbol{\theta}_{\mathrm{prop},i}}
\newcommand{\xDM}{\boldsymbol{x}_\mathrm{DM}}
\newcommand{\mDM}{m_\mathrm{DM}}

\renewcommand{\vec}[1]{\boldsymbol{#1}}

\arxivnumber{TTK-24-25}

\title{
$\thickbar{\mbox{\textsc{D}}}$\textsc{arkRayNet}: Emulation of cosmic-ray antideuteron fluxes from dark matter}

\author[a]{Jan Heisig,}
\author[b]{Michael Korsmeier,}
\author[a]{Michael Kr{\"a}mer,}
\author[a]{Kathrin~Nippel,}
\author[a]{and Lena Rathmann}

\affiliation[a]{Institute for Theoretical Particle Physics and Cosmology (TTK), RWTH Aachen University, Sommerfeldstr.~13, D-52056 Aachen, Germany}

\affiliation[b]{The Oskar Klein Centre for Cosmoparticle Physics, Department of Physics, Stockholm University, Alba Nova, 10691 Stockholm, Sweden}

\emailAdd{heisig@physik.rwth-aachen.de}
\emailAdd{mkraemer@physik.rwth-aachen.de}
\emailAdd{nippel@physik.rwth-aachen.de}
\emailAdd{rathmann@physik.rwth-aachen.de}

\abstract{Cosmic-ray antimatter, particularly low-energy antideuterons, serves as a sensitive probe of dark matter annihilating in our Galaxy. We study this smoking-gun signature and explore its complementarity with indirect dark matter searches using cosmic-ray antiprotons. To this end, we develop the neural network emulator \DRN, enabling a fast prediction of propagated antideuteron energy spectra for a wide range of annihilation channels and their combinations. We revisit the Monte Carlo simulation of antideuteron coalescence and cosmic-ray propagation, allowing us to explore the uncertainties of both processes. In particular, we take into account uncertainties from the $\Lambda_b$ production rate and consider two distinctly different propagation models. Requiring consistency with cosmic-ray antiproton limits, we find that AMS-02 shows sensitivity to a few windows of dark matter masses only, most prominently below 20\,GeV. This region can be probed independently by the upcoming GAPS experiment. The program package \DRN\ is available on \href{https://github.com/kathrinnp/DarkRayNet}{GitHub}.
}

\begin{document}

\maketitle
\section{Introduction}

The nature of dark matter (DM) is one of the most pressing questions in fundamental physics today, motivating a wide range of experimental programs aimed at detecting new elementary particles associated with this phenomenon (see e.g.~\cite{Bertone:2004pz,Arcadi:2017kky,Cirelli:2024ssz} for recent reviews). 
Indirect detection is a particularly important search strategy, as it probes the self-annihilating nature of DM and, hence, 
sheds light on the underlying production mechanism of DM in the early Universe. 
Provided that DM annihilates into matter and antimatter at equal rates, the searches for excess antimatter in the fluxes of antimatter in cosmic rays (CRs) offer promising prospects due to their low astrophysical background.
To date cosmic-ray antiprotons and positrons are well-established channels for DM searches, see~\cite{Cuoco:2017iax,Cholis:2019ejx,Kahlhoefer:2021sha,Calore:2022stf,Balan:2023lwg,DelaTorreLuque:2024ozf} for recent analyses of cosmic-ray antiproton fluxes. However, sizable uncertainties in the background predictions still make it hard to unambiguously associate an excess in the data to a signal of DM~\cite{Reinert:2017aga,Cuoco:2019kuu,Derome:2019jfs,Boudaud:2019efq,Heisig:2020nse,Heisig:2020jvs}.

For heavier anti-nuclei, such as antideuterons, the situation is different. While the background prediction is subject to similar uncertainties, its magnitude is several orders below the one of the targeted DM signal at sub-GeV energies,
since secondary antideuterons are kinematically suppressed in this regime.
Thus, the detection of antideuterons at low energies is a smoking gun signature of DM (see e.g.~\cite{Donato:1999gy,Donato:2008yx} and~\cite{Aramaki:2015pii,Korsmeier:2017xzj,Gomez-Coral:2018yuk,Perez:2019lxb, DeLaTorreLuque:2024htu} for an early and a recent account of this, respectively). While the experimental discovery of cosmic-ray antideuterons is still pending, the upcoming GAPS experiment~\cite{Aramaki:2015laa} will soon provide a significant step forward in testing their existence at low energies. Moreover, and intriguingly, over the past years AMS-02 has reported the detection of (a total of seven) tentative antideuteron events in preliminary analyses~\cite{Ting:2023CERNSeminar}.

In view of this expected experimental progress, it is a timely task to accurately predict the corresponding fluxes near Earth and their theoretical uncertainties, to allow for a sound interpretation of future observations. Our first major contribution to this effort is a reliable calculation of the predicted antideuteron source spectra from DM annihilation. This computation involves a Monte Carlo simulation of the DM annihilation process, the subsequent decay, showering, and hadronization of the produced Standard Model particles, and the formation of antideuteron bound states from the antiprotons and antineutrons produced as final states of the preceding processes (for a recently improved prediction of antiproton spectra see~Ref.~\cite{Arina:2023eic}). The bound state formation can be described by a simple coalescence model~\cite{Butler:1963pp,Schwarzschild:1963zz,Ibarra:2012cc,Fornengo:2013osa} or variations thereof (see Refs.~\cite{Gomez-Coral:2018yuk,Kachelriess:2019taq} for discussions). However, the model is subject to large uncertainties because the parameters of the coalescence model must be fixed by data, which are unfortunately sparse in the kinetic regime relevant for DM annihilation. Furthermore, the antideuteron spectra may be sensitive to poorly measured subprocesses. For example, Ref.~\cite{Winkler:2020ltd} has pointed out significant uncertainties in the antinuclear spectra from the poorly constrained production cross section of $\Lambda_b$ mesons, which can greatly affect the fluxes resulting from DM annihilation into $b\bar b$. Accordingly, we include the parameters of the coalescence model as nuisance parameters in our analysis.

Our second major contribution concerns the propagation of cosmic-ray antideuterons through our Galaxy and solar system. As a fully predictive theory of cosmic-ray propagation has not been established in the literature, the current state of the art is to empirically model cosmic-ray propagation as a diffusive process inferring the propagation model parameters from data. Several variations of propagation models have been discussed, whose guiding principles include theoretical motivation as well as minimization of the number of free parameters while being able to explain the data. Here we consider two of these models that have emerged from the literature~\cite{Genolini:2019ewc,Boudaud:2019efq,Weinrich:2020cmw,Korsmeier:2021brc}, the INJ.BRK (injection break) and DIFF.BRK (diffusion break) models, as introduced in~\cite{Korsmeier:2021bkw,Balan:2023lwg}. The propagation model parameters are subject to uncertainty because they are determined from data. Furthermore, their best-fit values could be affected by the presence of DM. For example, when extracting the diffusion coefficient from antiproton fluxes, the latter might contain a primary contribution from DM annihilation. Therefore, propagation parameters must be treated as nuisance parameters in the limit-setting procedure to allow profiling or marginalization.

In this paper, we address these challenges by revisiting the coalescence and propagation of antideuterons. By parameterizing the main uncertainties, we allow to estimate their impact on the antideuteron yield in upcoming experiments. Since both the Monte Carlo simulation of antideuteron formation from DM annihilation and the propagation of antideuterons are extremely CPU-intensive tasks, we use machine learning techniques to emulate both processes. This fast emulation is crucial to allow on-the-fly marginalization or profiling in the limit setting. 

Following the approach in \cite{Kahlhoefer:2021sha, Balan:2023lwg},\footnote{See Refs.~\cite{Lin:2019ljc,Tsai:2020vcx} for other applications of machine learning techniques to cosmic-ray propagation.} we use a recurrent neural network (RNN) to individually 
predict the antideuteron flux from DM annihilation and secondary emission. This network architecture is particularly well suited for emulating cosmic ray spectra because it is designed for sequential data and can thus naturally output the energy dependence of the antideuteron spectra. 
The network is trained on different cosmic ray propagation parameters in the DIFF.BRK and INJ.BRK models, and on a wide range of DM masses. The branching fractions into different SM final state particles serve as another input to the network, as well as the coalescence model parameters. 
This computational tool -- called \DRN\ -- has been made publicly available to the community.\footnote{\DRN\ is the most recent extension of the DarkRayNet code available on \url{https://github.com/kathrinnp/DarkRayNet}.}
It allows us to compute the marginalized flux near Earth and derive the projected sensitivities for AMS-02, the upcoming GAPS experiment and the future AMS-100 mission. We consider two different DM models -- annihilation into $b\bar b$ and the singlet scalar Higgs portal model -- and discuss the dependence of our results on the coalescence parameters.

The remainder of the paper is organized as follows. In Sec.~\ref{sec:dbar} we introduce the antideuteron production and propagation mechanisms and show our results for the antideuteron source spectra. The neural emulation of the propagated antideuteron spectra is described in Sec.~\ref{sec:DRN}, while our prediction for an antideuteron detection in AMS-02, GAPS, and AMS-100 is presented in Sec.~\ref{sec:results}. We conclude in Sec.~\ref{sec:concl}. Appendices~\ref{app:spatialsep} and \ref{app:LEPfit} contain details on the computation of the spatial separation of the antideuteron constituents and the calibration of the coalescence model, respectively. Appendix~\ref{app:LikeSen} describes how we marginalize the antideuteron flux over the propagation parameters. 

\section{Cosmic-ray antideuterons}
\label{sec:dbar}

Cosmic-ray antideuterons have not yet been discovered.\footnote{Preliminary results from AMS-02 suggest the detection of 7 antideuteron events~\cite{Ting:2023CERNSeminar}.} However, ongoing and upcoming experiments are increasing their sensitivity to reach progressively lower flux values. This increasing sensitivity raises the prospect of finally detecting antideuterons in CRs. 

\subsection{Production}
\label{sec:prod}

Similar to antiprotons which have been measured with good precision in the last decade, the production of CR antideuterons involves two possible processes. First, the standard astrophysical processes that result from the interaction of primary CRs with the gaseous matter in the Galactic disk. These processes lead to the production of secondary antideuterons. Second, the annihilation of DM particles within the CR diffusion halo can produce antideuterons. These resulting antideuterons exhibit a distinct energy spectrum peaking at lower energies, while secondary antideuterons are kinematically suppressed at low energies. This unique energy distribution makes CR antideuterons a promising channel to search and constrain DM models.

The common mechanism for both secondary antideuterons and antideuterons from DM annihilation is coalescence. Several models have been proposed that follow the same general principle: in the process of CR scattering or DM annihilation, antiprotons and antineutrons are produced. If the two particles are sufficiently close in momentum and position, they can fuse to form an antideuteron. 

The first calculation of secondary antideuterons in CRs was performed in the analytic coalescence model in Ref.~\cite{Chardonnet:1997dv}. In this reference, the cross sections for the production of antiprotons and antineutrons were taken from analytic parameterizations in a differential and a Lorentz invariant form. Assuming that the production is mostly prompt (i.e., that the spatial separation is negligible) and that the production of antiprotons and antineutrons is not correlated, it is possible to express the combined phase space density of the antiproton and antineutron in terms of integrals over the sum and over the difference of their momenta. Then an antideuteron is formed if, in the rest frame of the antideuteron, the momentum difference is smaller than a predefined threshold, the \emph{coalescence momentum} $p_c$. The coalescence momentum defined in this way is not known from first principles but must be determined by fitting the model to antideuteron production data from collider experiments.
Soon the approach was generalized to antideuteron production from DM, and in Ref.~\cite{Donato:1999gy} it was pointed out that low-energy antideuterons provide a unique signature to distinguish a DM signal from the astrophysical background. Since then, the analytical coalescence model has been used and refined in several works \cite{Duperray:2002pj, Duperray:2005si, Ibarra:2009tn, Brauninger:2009pe, Cirelli:2014qia,Carlson:2014ssa}.

The modeling of coalescence has been further improved by the use of Monte Carlo (MC) event generators. They allow the coalescence criterion to be considered on an event-by-event basis, taking into account both spatial information and the correlation of antineutron and antiproton production. In~\cite{Kadastik:2009ts, Ibarra:2013qt, Fornengo:2013osa, Herms:2016vop} and \cite{Dal:2014nda}, respectively, the event generators \textsc{Pythia}~\cite{Sjostrand:2014zea} and \textsc{Herwig}~\cite{Bellm:2015jjp} were used in this context. More recently, Ref.~\cite{Kachelriess:2020uoh} used \textsc{QGSJET}~\cite{Ostapchenko:2010vb,Ostapchenko:2013pia} and Ref.~\cite{Gomez-Coral:2018yuk} used EPOS-LHC~\cite{Pierog_2015} and Geant4's FTFP-BERT~\cite{AGOSTINELLI2003250} for the MC simulation. 

The correlation of antiproton and antineutron production has not been measured, and hence MC generators are not specifically tuned to it. It is therefore not surprising that different MC generators may require quite different values for the coalescence momentum to fit collider data. Furthermore, there remain spectral differences depending on the generator used~\cite{Dal:2014nda, Gomez-Coral:2018yuk}. 

A more elaborate model has been considered in ~\cite{Kachelriess:2019taq,Kachelriess:2020uoh,Kachelriess:2023jis}, which uses a quantum mechanical coalescence criterion based on the overlap of the wave functions of antiprotons and antineutrons. This approach follows the ideas outlined in \cite{Blum:2019suo}. However, the analysis of \cite{Kachelriess:2020uoh} uses 
\textsc{QGSJET}, so the problem of mismodeling of antiproton and antineutron correlations in the event generator remains. 

Recently, Ref.~\cite{Winkler:2020ltd} has pointed out that the displaced decay of $\Lambda_b$ baryons can significantly increase the DM-induced antinuclear fluxes, since the antiproton-antineutron pair is produced with very low relative momentum. In particular, the $\Lambda_b$ production cross section is subject to large uncertainties. In fact, \textsc{Pythia} falls short of the $b\to\Lambda_b$ transition ratio measured at LEP by a factor of about 3~\cite{Winkler:2020ltd}. This is an important effect in assessing the uncertainties of the coalescence model.

Finally, we note that thermal field theory models developed in the context of quark-gluon plasma physics can explain the production of antideuterons. However, the energy density relevant to the production of Galactic CR antineutrons is below the regime described by a quark-gluon plasma. 

In this work, we adopt the MC-based coalescence model using MadDM~3~\cite{Ambrogi:2018jqj} and \textsc{Pythia}~8.2~\cite{Sjostrand:2014zea} to simulate events. The former generates the hard process while the latter performs the showering and hadronization. For a given DM model parameter point, we generate a minimum of $10^8$ events. For each event, the momenta, displacements, and mother particles of all $\bar p$ and $\bar n$ in the event are considered. For each pairing of $\bar p$ and $\bar n$, we compute the relative momentum $\Delta^\mu= p^\mu_{\bar p} - p^\mu_{\bar n}$ and the spatial separation according to Appendix~\ref{app:spatialsep}. The event is considered to contain an antideuteron with momentum $p^\mu_{\bar d}=p^\mu_{\bar p} + p^\mu_{\bar n}$, if $\Delta^2 \le p_\text{c}^2$ for the pairing with the lowest $\Delta^2$ and if the corresponding spatial separation $\Delta r$ is smaller than 2\,fm (see e.g.~Ref.~\cite{Fornengo:2013osa}).\footnote{Note that usually only one pairing satisfies the $\Delta^2 \le p_\text{c}^2$ condition, so the order in which the momentum and spatial conditions are applied does not affect our results.} We use the coalescence momentum $p_c = 210 \,{}^{+27}_{-25}\,\,\big({}^{+48}_{-47}\big)\,$MeV which we derive from a fit to antideuteron production data from $e^+ e^-$ at the $Z$ resonance at LEP~\cite{ALEPH:2006qoi} as described in Appendix~\ref{app:LEPfit}. The quoted errors denote the 68\% (95\%) confidence level intervals. To account for the discrepancy in the $b\to\Lambda_b$ transition ratio when the result of \textsc{Pythia} is compared to observations from LEP, we introduce the rescaling of the $\Lambda_b$ production rate $r_{\Lambda_b}$ as a free parameter of our model and set its fiducial value to 3 as found in~\cite{Winkler:2020ltd}.
    
\subsection{Propagation}
\label{sec:prop}

Astrophysical sources are capable of accelerating CR nuclei to high energies. The CRs from these sources are called primary CRs. The most important primary nuclei are protons and helium. The CRs are then injected into our Galaxy, where they are affected by several processes. They are scattered by the magnetic turbulent fields, interact with the gas and photon fields in the Galaxy, may be transported by winds, and may be re-accelerated by magnetic waves. 

The most important process is the scattering by the turbulent magnetic fields in our Galaxy. Effectively, this propagation process can be described by a diffusion model. The diffusion coefficient depends on the rigidity $R$ of the particle. We model it by a double-broken power law with smooth breaks:
\begin{eqnarray}
    \label{eqn:diffusion_coefficient}
     D(R) &\propto& \beta R^{\delta_l}
 	  \left[ 1 + \left(\frac{R}{R_{D,0}}\right)^{\frac{1}{s_{D,0}}} \right]^{s_{D,0}\,( \delta - \delta_l) } 
 	\,\left[ 1 + \left(\frac{R}{R_{D,1}}\right)^{\frac{1}{s_{D,1}}} \right]^{s_{D,1}\,( \delta_h - \delta) },
\end{eqnarray}
where $R_{D,0}$ and $R_{D,1}$ are the positions of the two breaks smoothed by $s_{D,0}$ and $s_{D,1}$, respectively. The parameters $\delta_l$, $\delta$, and $\delta_h$ denote the spectral indices below, between, and above the breaks, respectively. Finally, $\beta$ is the velocity of the particle (in units of $c$).
The diffusion coefficient is normalized such that $D(R\!=\!4\,\text{GV})=D_0$, where $D_0$ is another parameter of the propagation model.

The injection of the primary nuclei is given by:
\begin{equation}
 \label{eqn:energy_spectrum}
 q_i(R) = \left(\frac{R}{R_0}\right)^{-\gamma_{1}} \left( \frac{R_0^{1/s} + R^{1/s}}{2 R_0^{1/s}} \right)^{-s(\gamma_{2} - \gamma_{1})}
\end{equation}
where $\gamma_{1}$ and $\gamma_{2}$, respectively, are the spectral indices\footnote{For protons, we use separate spectral indices $\gamma_{1,p}$ and $\gamma_{2,p}$ to account for the observed difference in the slopes in the cosmic ray fluxes of the p and He spectra.} below and above the smooth break at position $R_0$ with the smoothing parameter $s$.
On the other hand, secondary CRs are produced by the fragmentation of primary CRs. The typical example is B, which is not produced in the stellar cycle and therefore mostly comes from the fragmentation of primary C, N, and O. 

We model all these phenomena by considering a chain of coupled diffusion equations solved with the numerical code \textsc{Galprop} version 56~\cite{Strong:1998fr} and \textsc{Galtoollibs} 885. We have made several modifications to the code. First, we have implemented the source terms for secondary and tertiary antideuterons using the analytic coalescence model as described in \cite{Kadastik:2009ts}. Second, we have added the source term for DM antideuterons by interpolating the tabulated annihilation spectra we have computed. Third, we have included the inelastic and tertiary cross section for antideuterons. Furthermore, regarding the propagation model, we implemented a smoothly broken power law for the diffusion coefficient and for the injection spectra as described above. Finally, we have improved the antiproton production cross sections according to the results of Ref.~\cite{Korsmeier:2018gcy}.

In our analysis, we explore two different models of CR propagation, denoted INJ.BRK and DIFF.BRK. These models correspond to those considered in Ref.~\cite{Balan:2023lwg}, to which we refer for further details. The models come with the following sets of assumptions and free parameters:

\begin{enumerate}
    \item[I.] In the INJ.BRK (injection break) model, we use a broken power law for the injection spectra of the primary CRs following Eq.~\eqref{eqn:energy_spectrum}. In our fit, we allow for different slopes of the injection spectra for proton and He  to account for the observed difference in the respective CR fluxes. The origin of this difference is subject to ongoing research and ranges from different source populations~\cite{1212.0381,1610.06187,1905.06699} to a $Z/A$-dependence of the efficiency of Fermi shock acceleration~\cite{1704.08252,1803.00428}. The diffusion coefficient is modeled as a single broken power law, \emph{i.e.}~setting $\delta_l=\delta$ in Eq.~\eqref{eqn:diffusion_coefficient}, with the break position, $R_{D,1}$, around 300 GV\@. We take into account reacceleration and convection parametrized by $v_{\rm A}$ and $v_{0,c}$, respectively. 
This model has been extensively studied in the literature~\cite{astro-ph/9807150, astro-ph/0101231,0909.4548,1602.02243,1607.06093,2006.01337,2102.13238}. 
\item[II.]
In the DIFF.BRK (diffusion break) model, we assume a single power law for the injection spectrum of the primary CRs (\emph{i.e.}~$\gamma_{1}=\gamma_{2}$ in Eq.~\eqref{eqn:energy_spectrum}) and no reacceleration, but use the double-broken power law of Eq.~\eqref{eqn:diffusion_coefficient} restoring $\delta_l$, $R_{D,0}$, and $s_{D,0}$ as free parameters of the model. This propagation model has been studied e.g.~in \cite{astro-ph/9807150} and has recently been tested against AMS-02 data in Refs.~\cite{1904.05899,Weinrich:2020cmw, Korsmeier:2021brc}. The additional break has been attributed to a damping effect of turbulences at low energies due to the interaction of CRs with turbulent magnetic fields~\cite{astro-ph/0510335}.
\end{enumerate}

For both models, we assume a half-height of the diffusion halo of $z_h=\unit[4]{kpc}$, which roughly corresponds to the lower bound from beryllium data from AMS-02 \cite{1910.04113, 2004.00441, 2102.13238, Korsmeier:2021brc, 2203.07265}.
Note that a precise determination of the halo size is currently hindered by significant systematic uncertainties in the fragmentation cross sections of secondaries~\cite{1803.04686,1910.04113, Korsmeier:2021brc} allowing for larger values for the half-height. This has a direct impact on the predicted cosmic-ray fluxes resulting in sizeable uncertainties, see e.g.~\cite{Genolini:2021doh}. However, we have conservatively chosen the above small value following the reasoning of~\cite{Balan:2023lwg}.
Finally, solar modulation is modeled by the force-field approximation, which allows for slightly different solar modulation potentials for antiprotons compared to protons and He \cite{AMS:2018avs}.

In the case of AMS-02 measurements, we can treat the force-field potential as a nuisance parameter to be marginalized over. This is done by including the potential in a parameter scan that is used to collect appropriate training data for the emulator, as discussed in section~\ref{sec:DRN}. We thus obtain the potential for each parameter point that we later marginalize over. 
For future experiments, we need to fix the potential for all data points. In the case of GAPS, the solar modulation potential for antideuterons detected by GAPS is set to $\phi = 700$\,MV, as GAPS is scheduled to operate in the winter of 2024/2025\footnote{\url{https://gaps1.astro.ucla.edu/gaps/}} when the solar potential will be near its maximum~\cite{2019JGRA..124.2367K}. As the solar potential for AMS-100 cannot yet be predicted, we set the value to $\phi = 600$\,MV which is roughly the mean of all measured solar potential values.

\subsection{Dark Matter}
\label{sec:DMsource}

Besides secondary and tertiary antideuterons from astrophysical sources, we consider primary antideuterons from DM annihilation. 
The corresponding source term is
\begin{eqnarray}
  \label{eq:pbar_DM_source_term}
  q_{\bar{d}}^{(\mathrm{DM})}(\bm{x}, E_\mathrm{kin}) =
      \frac{1}{2} \left( \frac{\rho(\bm{x})}{m_\mathrm{DM}}\right)^2  
      \left\langle \sigma v \right\rangle_\text{tot}\sum_f {\cal B}_f \frac{\diff N^f_{\bar{d}}}{\diff E_\mathrm{kin}} \; ,
\end{eqnarray}
where $m_\text{DM}$ and $\rho(\bm{x})$ denote the DM mass and energy density profile, respectively, and  $\langle \sigma v \rangle_\text{tot}$ is the velocity-averaged annihilation cross section times velocity.  The sum runs over all contributing  annihilation channels $f$ with relative contribution ${\cal B}_f = \langle \sigma v \rangle_f/ \langle \sigma v \rangle_\text{tot}$ and energy spectrum $\diff N^f_{\bar{d}}/\diff E_\mathrm{kin}$ at the source. 
For the energy density profile, we assume a NFW radial profile \cite{Navarro_1996} with a scale radius of $r_h = 20\,$kpc and a local DM density at solar position of 0.43 GeV/$\mathrm{cm}^3$. In this work, we consider $m_\text{DM}$ and ${\cal B}_f$ as free parameters of the DM model.

Using the coalescence model and following the methodology described in Sec.~\ref{sec:prod}, we compute the energy source spectra $\diff N^f_{\bar{d}}/\diff E_\mathrm{kin}$ for DM between 2\,GeV and 10\,TeV\footnote{We use a grid of 37 mass points for the entire range.} and for a wide range of annihilation channels, specifically, $q\bar q$, $c\bar c$, $b\bar b$, $t\bar t$, $gg$, $hh$, $W^+W^{-(*)}$, and $ZZ^{(*)}$, where the asterisk denotes an off-shell gauge boson relevant for DM masses below the respective pair-production threshold. Note that $q$ corresponds to the three light quark flavors, $u,d,s$, whose annihilation spectra are virtually identical.

\begin{figure}
   \centering
    \includegraphics[width = 0.9\textwidth]{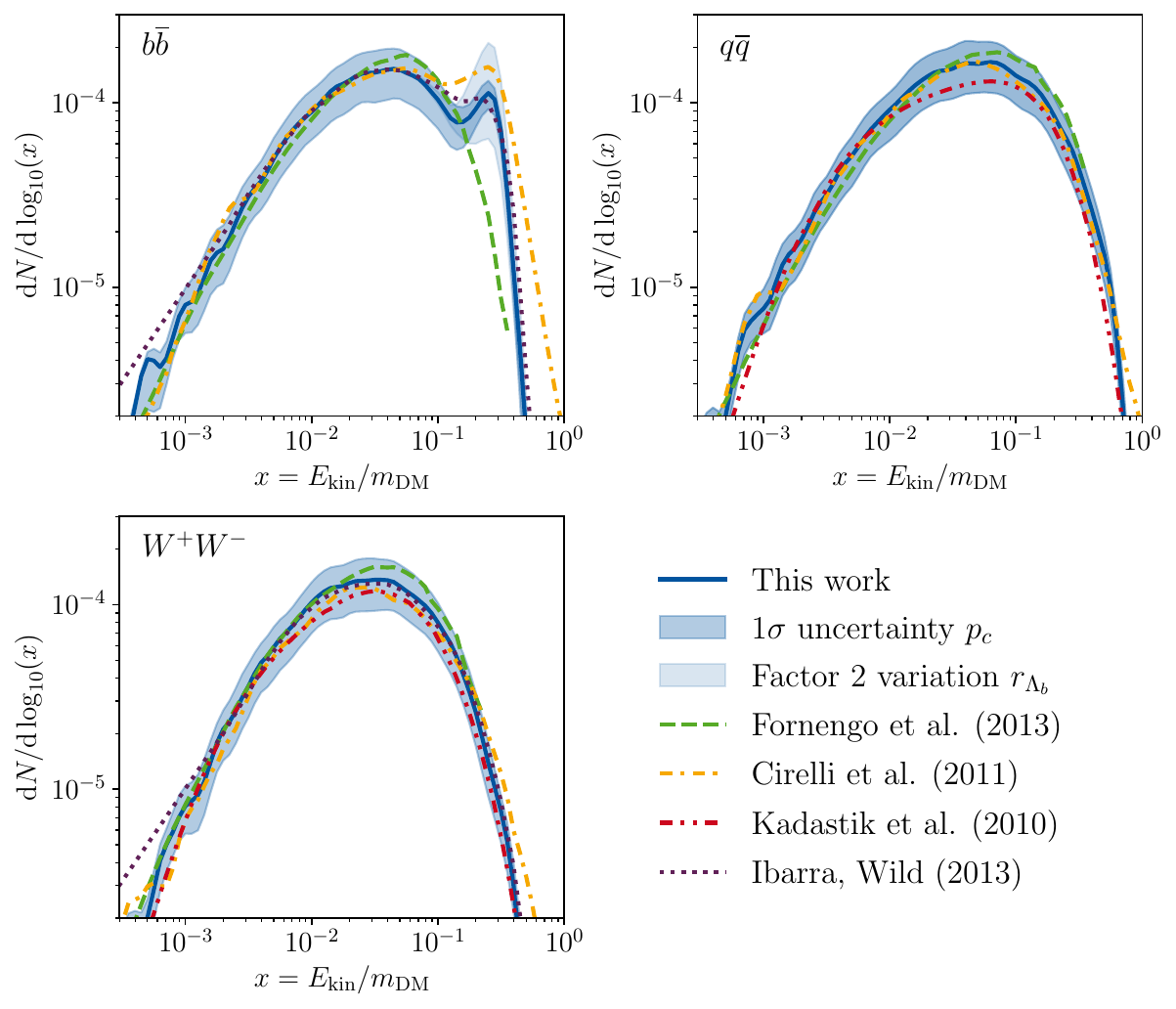}
    \caption{Comparison of injection spectra for different annihilation channels for $m_{\mathrm{DM}} = 100\,$GeV. The panels show annihilation via the $b\overline{b}$ \textit{(top left)}, $q\overline{q}$ \textit{(top right)} and $W^+W^-$ \textit{(bottom left)} channels. The darker shaded blue region shows the $1\sigma$ uncertainty range of the coalescence parameter $p_c$. The lighter shaded blue region in the $b\overline{b}$ spectrum shows the region of a factor 2 increase and decrease in the production rate of $\Lambda_b$ baryons from $b$ quarks.}
    \label{fig:SourceSpectra}
\end{figure}

In Fig.~\ref{fig:SourceSpectra} we show our results for the source spectra using the best-fit value for $p_c$ from Eq.~\eqref{eq:p_c_value} for the three exemplary annihilation channels $b\bar b$, $q\bar q$, and $W^+W^-$, as well as the $1\sigma$ uncertainty range for $p_c$. We also show the variation of the scaling of the $\Lambda_b$ production rate $r_{\Lambda_b}$ by a factor of 2 in the $b\bar b$ spectrum. For comparison, we present previous results from the literature, specifically from Cirelli \emph{et al.}~\cite{PPPC4DMID}, Ibarra and Wild~\cite{Ibarra:2012cc}, Fornengo \emph{et al.}~\cite{Fornengo:2013osa}, and Kadastik \emph{et al.}~\cite{Kadastik:2009ts}. 

There are a few differences in the generation of the injection spectra computed in the references shown in the figure, which lead to discrepancies in the spectra. The first difference is the use of different versions of the shower algorithm \textsc{Pythia}. The different spectra shown here were produced using \textsc{Pythia}~8.1 (except for Ref.~\cite{Fornengo:2013osa}~which uses \textsc{Pythia}~6.4), while we use the newer version \textsc{Pythia}~8.2. In particular, the use of an older \textsc{Pythia} version in Ref.~\cite{Fornengo:2013osa} may explain why these spectra are systematically higher than all other spectra. The differences in using different shower algorithms have also recently been discussed in \cite{Arina:2023eic}. 

The biggest discrepancies can be seen in the source spectra for DM annihilation into $b \bar{b}$ between $x=0.1$ and 1, where some spectra display a bump while others do not. This difference is due to different treatment of antideuterons produced by long-lived particles. When DM annihilates and long-lived particles are produced during hardronization and decay of the produced SM particles, the decay of the long-lived particle will take place at a displaced vertex. The distance of antiprotons or antineutrons produced at the displaced vertex to the particles produced in prompt decays is too large for the formation of an antideuteron because of the finite range of the nuclear forces. Only if the distance between the antiproton and the antineutron is below a few fm can they bind and form an antideuteron \cite{Ibarra:2012cc}. This formation can only happen at the initial vertex or through the off-vertex production of a $\bar{\Lambda}_b$ particle as described in \cite{Winkler:2020ltd}. Because of the rest mass of 5.6$\,$GeV, the resulting multi-nucleon states have very small relative momenta which is necessary for the production of antideuterons. Therefore, if particles produced at displaced vertices are ignored then there is no bump between $x=0.1$ and 1. In contrast to that, if the spatial separation of the produced antiprotons and antineutrons is not checked, then the amount of antideuterons is overestimated and the bump is too large. In our approach, we take the production of antideuterons from $\bar{\Lambda}_b$ particles into account and check whether the spatial separation of the antiproton and antineutron is below 2$\,$fm.

Because the coalescence momentum has to be determined from experiments and the value varies depending on the experiment and the Monte Carlo generator used, the values used in the various analyses are not the same in the spectra we present. The spectra are roughly proportional to $p_\text{c}^3$. This leads to an overall rescaling of the spectra of $(p_\text{c}/p_\text{c,0})^3$ if the coalescence momentum $p_\text{c,0}^3$ is used for the computation of the spectra. This can explain off-sets when comparing different spectra. 

Also note that the light quark content, denoted by $q$, is different in the different analyses. Some include only the lightest quark $u$,  while others also include quarks up to charm quarks. As these particles are still very light compared to the DM mass, the difference in the spectra should be negligible.

\section{Neural Emulation of $\dbar$ with the \DRN} \label{sec:DRN}

\begin{figure}
   \centering
	\begin{subfigure}{0.49\textwidth}
        \centering
		\includegraphics[height=0.33\textheight,trim={0.0cm 2cm 0.0cm 0.0cm},clip]{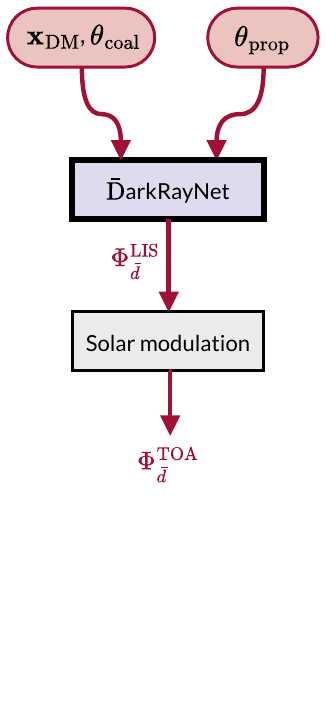}
        \caption{Single flux}
	\end{subfigure}
     \begin{subfigure}{0.49\textwidth}
        \centering
		\includegraphics[height=0.4\textheight]{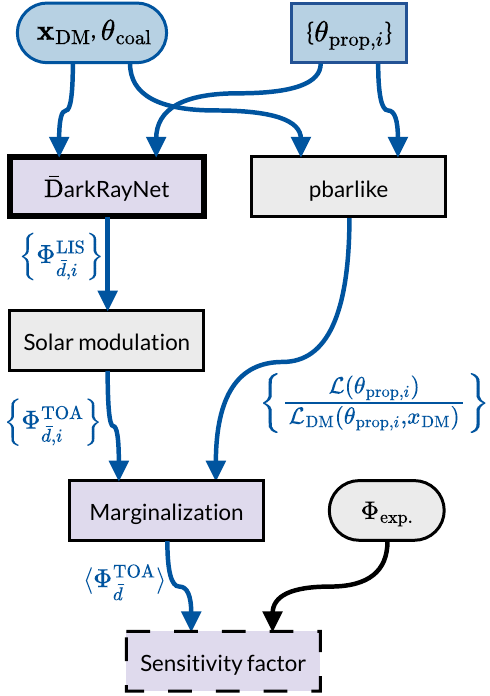}
        \caption{Marginalized flux}
        \label{fig:toolchain2}
	\end{subfigure}
    \caption{Tool-chain for obtaining the antideuteron flux and the sensitivity factor. The rounded boxes show the input and the rectangular boxes correspond to different steps in our analysis. In purple we show our new contributions.}
    \label{fig:toolchain}
\end{figure}

In this section, we describe the neural network setup and tool-chain that allows us to emulate propagated $\dbar$ spectra and enables fast predictions for projected experimental sensitivities. The central tool is the neural emulator \DRN, which computes the local interstellar antideuteron flux $\Phi^{\mathrm{LIS}}_{\bar{d}}$ for a given set of DM parameters $\vec{x}_{\mathrm{DM}}=\{\mDM,{\cal B}_f\}$, coalescence parameters $\vec{\theta}_{\mathrm{coal}}=\{p_c,r_{\Lambda_b}\}$, and propagation parameters $\vec{\theta}_{\mathrm{prop}}$. To obtain the top-of-atmosphere flux, $\Phi_{\bar d}^\text{TOA}$, we link the \DRN\ to a solar-modulation module modeling the propagation of antideuteron through the heliosphere in the force-field approximation as described in Sec.~\ref{sec:prop}.
The tool-chain is schematically shown in the left diagram in Fig.~\ref{fig:toolchain}. We describe the training sample and neural network used in the \DRN\ in Sec.~\ref{sec:train} and \ref{sec:DRNsub}, respectively.

To enable the inclusion of propagation uncertainties in the computation of projected sensitivities (or exclusion limits) via marginalization, we provide the tool-chain displayed in the right diagram in Fig.~\ref{fig:toolchain}, which contains additional modules. 
These include a set of propagation parameters $\{\theta_{\mathrm{prop},i}\}$ representing their posterior distribution under the assumption of no existing DM signal. Furthermore, the antiproton likelihood calculator \textsf{pbarlike}~\cite{Balan:2023lwg}, and a marginalization routine that provides the marginalized flux $\langle \Phi \rangle_{\overline{d}}$ are added to the pipeline. 
These functionalities are detailed in Sec.~\ref{sec:Marg}.
We exemplify the use of our tool-chain for computing a sensitivity factor for several experiments.

\subsection{Training sample}
\label{sec:train}

For the generation of the training sample, we randomly scan the input parameters $\vec{x}_{\mathrm{DM}}$ and $\vec{\theta}_{\mathrm{coal}}$ on a hypercube within their region of interest. The DM mass, $\mDM$, the rescaling of the $\Lambda_b$ production rate, $r_{\Lambda_b}$, and the relative contributions of annihilation channels, ${\cal B}_f$, are sampled logarithmically in the range $[2,10^4]$\,GeV, $[0.3,20]$, and  $[10^{-5},1]$, respectively. 
For the latter, we consider the eight channels $q\bar q$, $c\bar c$, $b\bar b$, $t\bar t$, $gg$, $hh$, $W^+W^{-(*)}$, and $ZZ^{(*)}$ and renormalize them to yield $\sum_f{\cal B}_f=1$ after random sampling.
The coalescence momentum $p_c$ is sampled with a linear prior in the range (148--300)\,MeV, which includes the 95\% confidence level interval.

We sample the remaining input parameters, the propagation parameters $\vec{\theta}_{\mathrm{prop}}$, from a large set of samples probed in a fit of $\pbar$ spectra from secondary emission, as well as proton and Helium spectra to AMS-02 data~\cite{Aguilar:2021tos} using a nested sampling approach. This procedure has been adopted from~\cite{Balan:2023lwg} and we refer to that reference for more details and information on the corresponding priors. 
We use the error correlations of the AMS-02 measurements from~\cite{Heisig:2020nse, Boudaud:2019efq}.
The fit gives us a broad range of propagation parameter sets, thoroughly covering the physically suitable parameter ranges. The physical interpretation of the parameters is described in section~\ref{sec:prop}. 

For a given $\vec{x}_{\mathrm{DM}}$ and $\vec{\theta}_{\mathrm{coal}}$, the injection spectra are obtained from an interpolation of pre-computed tables following the prescription laid out in Sec.~\ref{sec:DMsource}. The spectra are then fed into \textsc{Galprop} to account for propagation effects according to Sec.~\ref{sec:prop}.\footnote{
To save computational resources, for the channels $W^+W^{-(*)},ZZ^{(*)}$, we emulate the propagation by a separate neural network~\cite{GalTrained} we trained on the above-mentioned sample.}
We also compute the corresponding spectra for secondary antideuterons. The set of input parameters and corresponding propagated spectra from DM annihilation and from secondary emission constitute the training sample for our neural network described in the next section.

\subsection{Network setup and performance}
\label{sec:DRNsub}

The setup of the emulation networks for antideuterons fluxes is similar to the emulator for predicting $\bar{p}$ in \cite{Kahlhoefer:2021sha}. We set up individual networks for antideuterons from secondary emission and DM annihilation, in order to be able to predict them separately and depending on the relevant physical parameters. This allows us to rescale the annihilation cross section of our DM models a posteriori. We use the API \textsc{Keras}~\cite{chollet2015keras} which is based on \textsc{Tensorflow}~\cite{DNN:tensorflow} for the network setup.

\begin{figure}
   \centering
    \includegraphics[width = 0.9\textwidth]{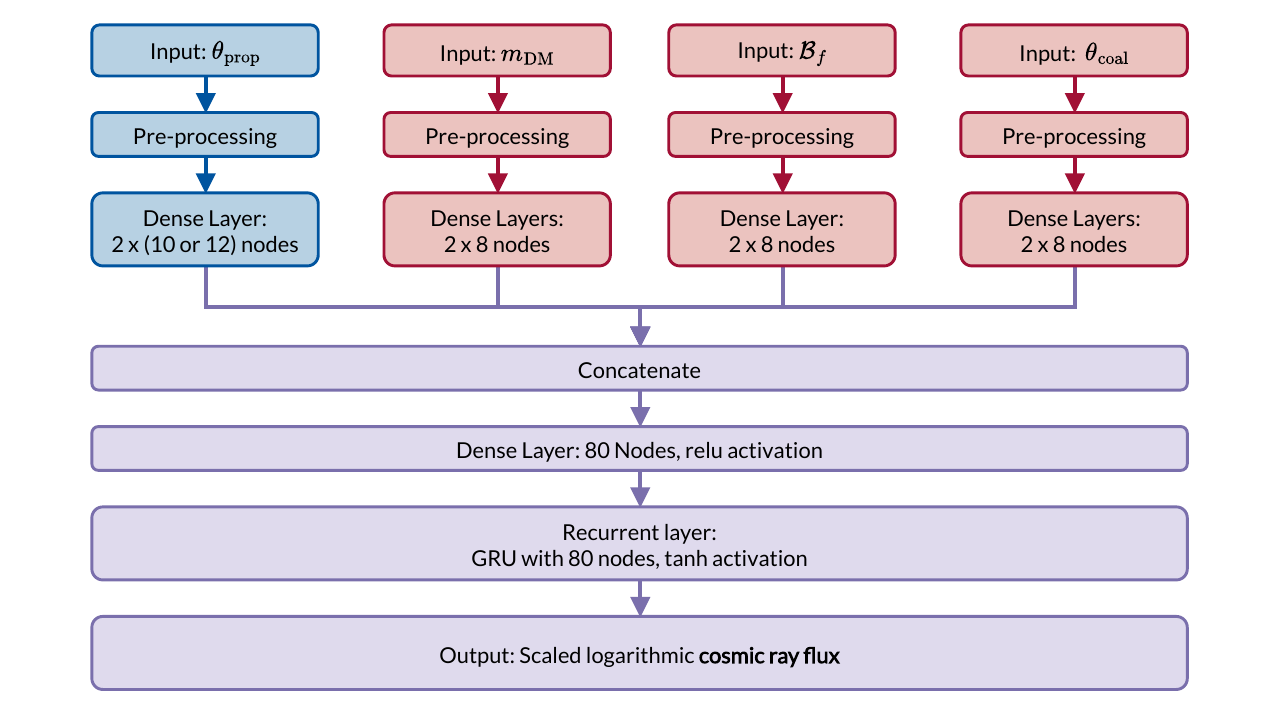}
    \caption{Sketch of the network setup to predict the flux of antideuterons from DM annihilation. The number of nodes for the propagation parameter input depends on the propagation model (12 for INJ.BRK and 10 for DIFF.BRK).}
    \label{fig:network_architecture}
\end{figure}

The structure of the network for antideuterons from DM annihilation, including the architecture hyperparameters, is shown in Fig.~\ref{fig:network_architecture}. We pre-process the four (sets of) parameters, $\vec\theta_\text{prop}$, $\mDM$, ${\cal B}_f$, and $\vec\theta_\text{coal}$  separately each with an independent dense network. These processed inputs are then combined and used as the input to a recurrent network consisting of a dense layer, a recurrent layer, and another dense layer as the final layer. The recurrent layer is chosen to improve performance by accounting for the correlation between energy bins, as discussed in~\cite{Kahlhoefer:2021sha}. The output of the network is the propagated antideuteron flux.

\begin{figure}
   \centering
    \includegraphics[width = \textwidth]{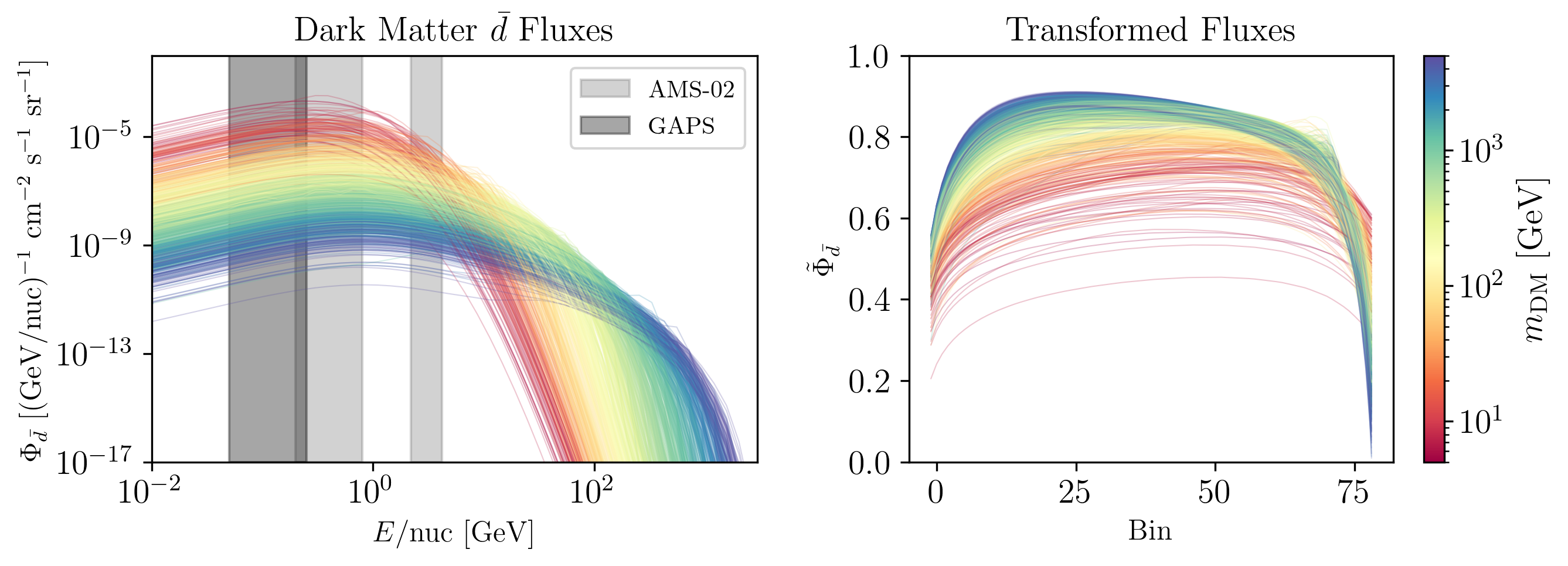}
    \caption{Examples of antideuteron fluxes without solar modulation in the INJ.BRK model (left) and their transformation (right). The transformed fluxes are used as the network input. The color sale indicates the DM mass. The shaded gray regions indicate the energy ranges of GAPS and AMS-02 in which the experiments are sensitive to antideuterons.}
    \label{fig:flux_trafo}
\end{figure}

To improve the performance of the network, we transform the input DM parameters to be in the range [0,1] and express the propagation parameters in terms of a distribution with a mean of 0 and a standard deviation of 1. We transform the propagated antideuteron fluxes onto 80 bins so that the flux is in the range [0,1]. The rescaled flux is calculated using
\begin{equation}
    \Tilde{\Phi}_{\bar{d}} = 1 + \frac{1}{10} \, \mathrm{log} \left(
    \frac{\Phi_{\bar{d}}}{\mathrm{GeV}^{-1}\,\mathrm{cm}^{-2}\, \mathrm{s}^{-1} \,\mathrm{sr}^{-1}}  \left(\frac{m_\mathrm{DM}}{\mathrm{GeV}}\right)^2 \left(\frac{p_{c,0}}{p_{c}}\right)^3  \frac{E/\mathrm{n}}{\mathrm{GeV}} \right)\,,
\end{equation}
where $p_{c,0}$ is our best-fit coalescence momentum (see appendix \ref{app:LEPfit}) and $E/$n is the kinetic energy per nucleon. This transformation conveniently maps almost all the fluxes onto the range [0,1]. The procedure can easily be inverted to obtain the correct flux from the output of the network. For very small fluxes, the transformation can map parts of the spectrum to values smaller than zero. These parts are then set to zero. The induced inaccuracy is, however, irrelevant for the results obtained, as it concerns fluxes well below any (future) sensitivity. The transformation for some randomly selected fluxes in the INJ.BRK model is shown in Fig.~\ref{fig:flux_trafo}.

Training is done using the \textsc{Adam} optimizer. We use a mean absolute error to evaluate the difference between the predicted and true fluxes. Starting with an initial learning rate of $10^{-2}$, we use a \texttt{ReduceLROnPlateau} schedule for training. Training is stopped when the validation loss does not decrease for 20 epochs using an \texttt{EarlyStopping} callback. 
As there are no new physical parameters in the case of antideuterons from secondary emission compared to secondary antiprotons, we adopt the same architecture as in~\cite{Kahlhoefer:2021sha}. Here we use the Adam optimizer with a mean squared error loss.

\begin{figure}
	\centering
	\begin{subfigure}[t]{0.49\textwidth}
        \centering
		\includegraphics[width=\textwidth]{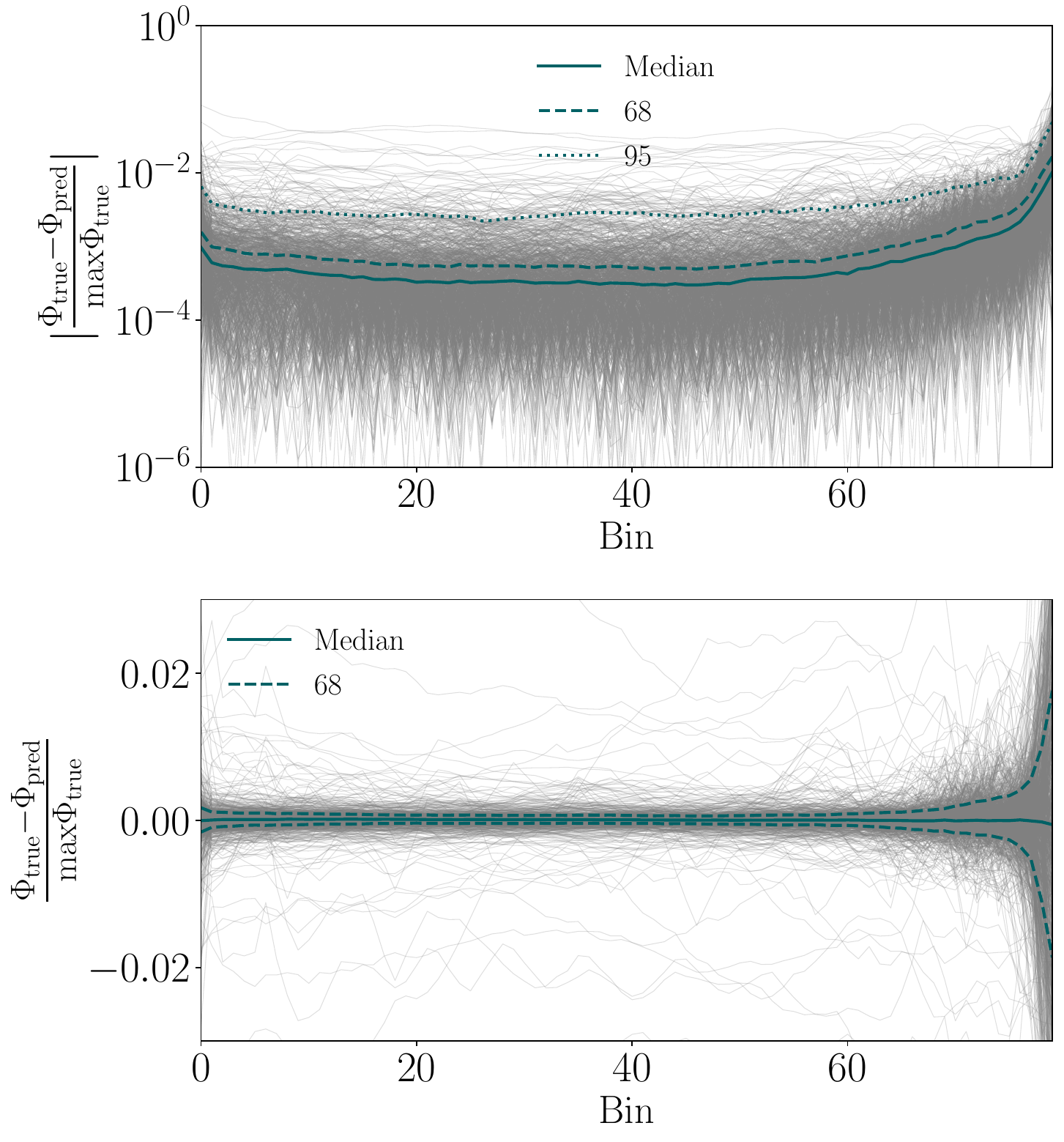}
        \caption{INJ.BRK}
	\end{subfigure}
	\hfill
        \begin{subfigure}[t]{0.49\textwidth}
        \centering
		\includegraphics[width=\textwidth]{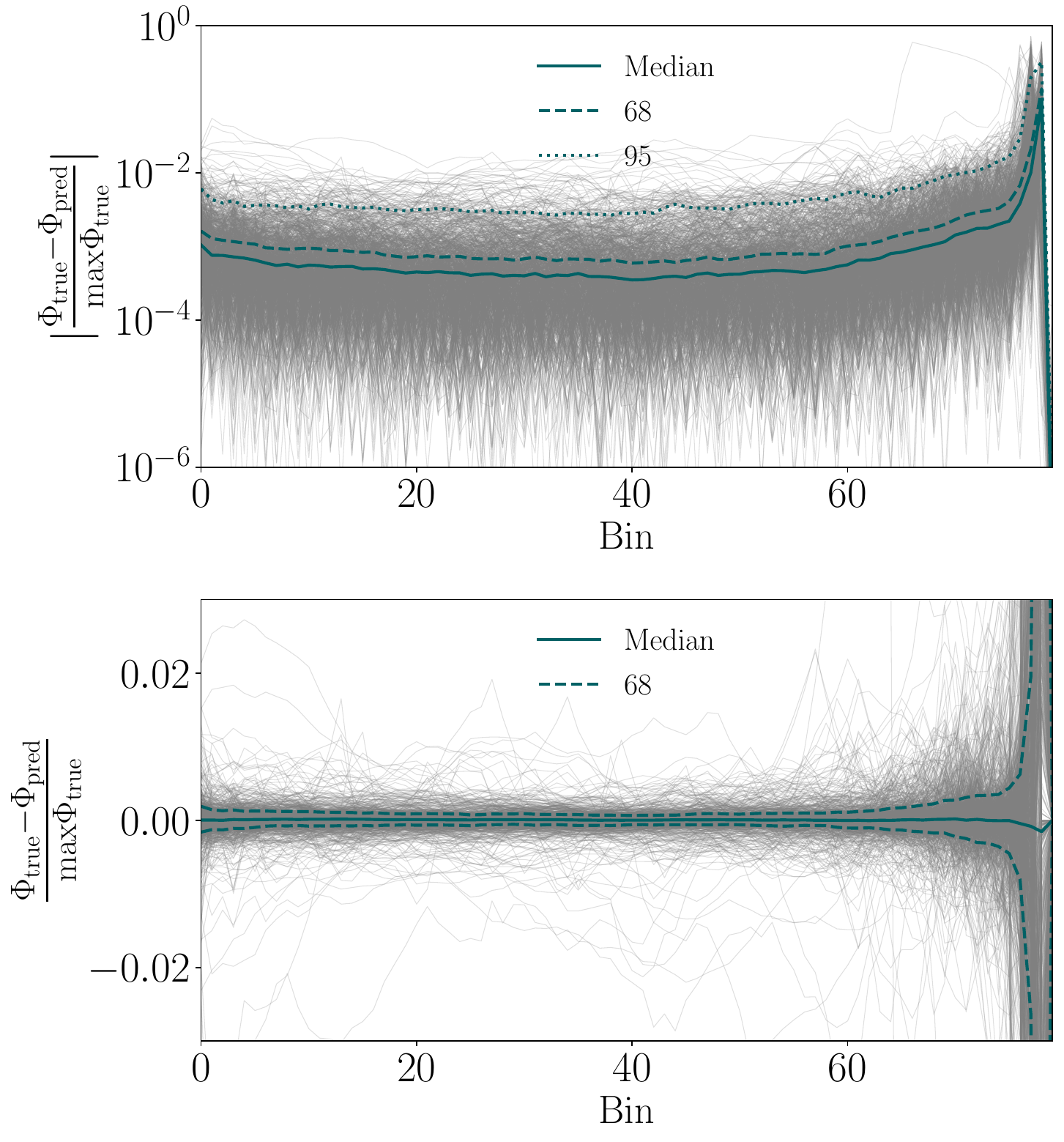}
        \caption{DIFF.BRK}
	\end{subfigure}
	\caption{Comparison of the predicted transformed flux and the true transformed flux in both propagation models for 1000 randomly selected fluxes. The solid blue line shows the median relative deviation of all spectra while the dashed and dotted lines show the 68 and 95 percentile of the deviations respectively.}
    \label{fig:network_performance}
\end{figure}

Once our network is fully trained on a large data set we can evaluate the accuracy of the network. To estimate its performance, we compare the predicted transformed flux of the network to the actual transformed flux using data that has not been used by the network during the training process. This comparison is shown in Fig.~\ref{fig:network_performance}, where we compute the difference between the prediction and the truth normalized to the maximum of the true flux in each bin for the INJ.BRK and DIFF.BRK models. The top panel shows the absolute value of the difference, while the bottom panel shows the total difference. 

In both cases, the network performs very well. In the majority of the bins, 68\,\% of all transformed fluxes have a relative difference of at most $6 \times 10^{-4}$. Only the last few bins display a larger uncertainty. These bins translate to energy bins in which the flux is very small and therefore hard to learn and also not important. The uncertainty in the transformed spectra translates into a relative error of $\mathcal{O}(10^{-2})$ in the actual flux. Therefore we conclude that the accuracy of the network is sufficient and it can be used to accurately predict the antideuteron flux.

\subsection{Marginalized spectra and sensitivity factor}
\label{sec:Marg}

In this subsection, we describe the marginalization over propagation parameters and computation of projected sensitivities, \emph{i.e.}~the tool-chain depicted in the diagram in Fig.~\ref{fig:toolchain2}.
To compute the marginalized flux, we utilize the posterior sample of the propagation parameters, $\{\vec\theta_{\mathrm{prop},i}\}$. This sample is obtained from the nested sampling fit, which is also used to obtain the training sample, see Sec.~\ref{sec:train} for details. 
For a given set of DM and coalescence parameters, we run \DRN\ and the solar modulation module for each $\vec\theta_{\mathrm{prop},i}$ of the sample and obtain the respective flux, $\Phi_{\bar d, i}^\text{TOA}$. Additionally, we compute the likelihood ratio $\mathcal{L}_\mathrm{DM} (\tpropi, \xDM)/\mathcal{L} (\tpropi)$ from AMS-02 flux measurements (see Sec.~\ref{sec:train}) for each sample point using \textsf{pbarlike}~\cite{Balan:2023lwg}. The marginalized flux, $\langle \Phi_{\bar d}^\text{TOA}\rangle$, is then obtained by the sum $\sum_i\Phi_{\bar d, i}^\text{TOA} \mathcal{L}_\mathrm{DM} (\tpropi, \xDM)/\mathcal{L} (\tpropi)$ normalized by the evidence ratio, see Appendix~\ref{app:LikeSen} for details. Marginalization can be performed within each of the two propagation models considered.

\begin{figure}
   \centering
    \includegraphics[scale=0.5]{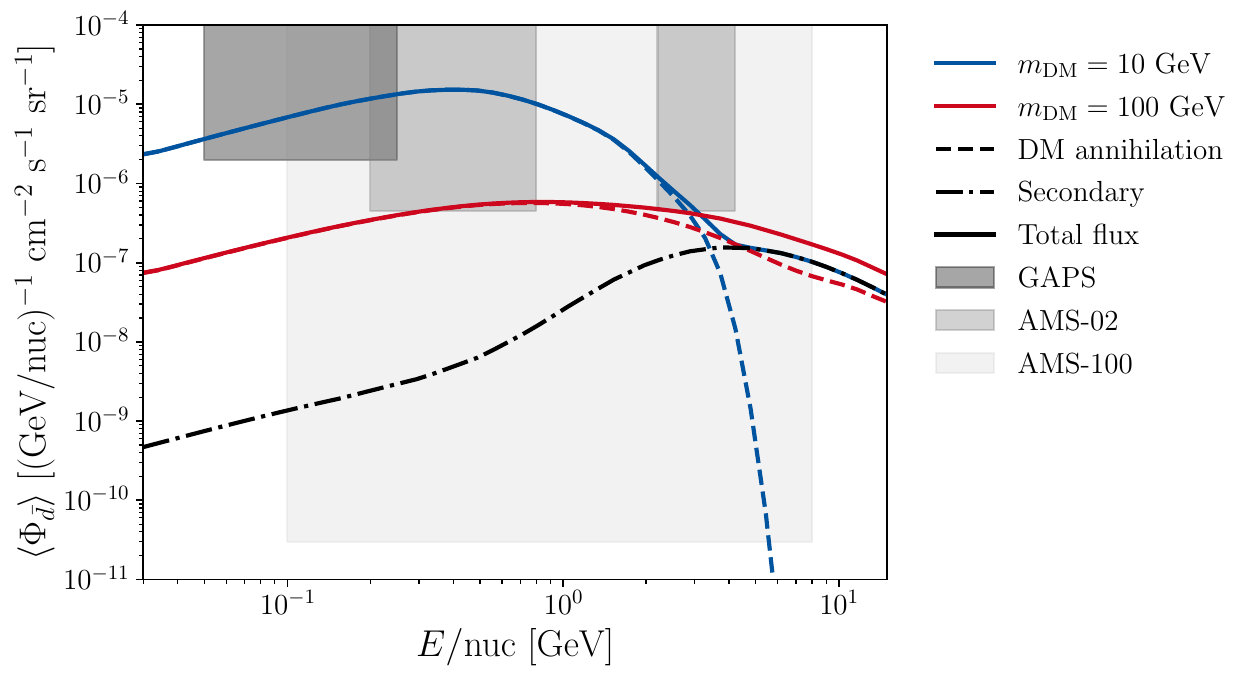}
    \caption{Two flux examples for $b \bar{b}$ annihilation in the INJ.BRK model for DM masses of 10 and 100 GeV. The fluxes are marginalized over all propagation parameters and the solar potential of GAPS was assumed. The fluxes are calculated assuming a thermal cross section of $\langle \sigma v \rangle = 3 \times 10^{-26}\, \mathrm{cm}^3 \mathrm{s}^{-1}$. Secondary antideuterons are shown with the dashed-dotted line. The shaded regions indicate the antideuteron flux sensitivities of GAPS, AMS-02 and AMS-100.}
    \label{fig:propspectbb}
\end{figure}

In Fig.~\ref{fig:propspectbb} we show the marginalized flux for two exemplary points in DM parameter space considering the INJ.BRK propagation model and nominal values for $\vec\theta_\text{coal}$. The blue and red lines show the antideuteron flux for DM with a mass of $m_{\mathrm{DM}} = 10\,$GeV and $m_{\mathrm{DM}} = 100\,$GeV respectively, assuming the thermal cross section, $\langle \sigma v \rangle = 3 \times 10^{-26}\, \mathrm{cm}^3 \mathrm{s}^{-1}$, and DM annihilation into $b \bar{b}$. The dashed and dot-dashed curves also show the individual contributions from DM and secondary emission. Note that the latter is only plotted for $m_{\mathrm{DM}} = 10\,$GeV but is very similar for the other mass (and is hence not shown to reduce clutter). Additionally, we show the experimental antideuteron flux sensitivities for GAPS, AMS-02, and AMS-100~\cite{Aramaki:2015laa,Choutko2008,Schael_2019}. They are listed in Table~\ref{tab:sensitivities}.

\begin{table}[h]
\begin{center}
\begin{tabular}{ c | c | c | c}
\multirow{2}{*}{Experiment} & Energy range & 
$\Phi_{\sensitivity, E_\exper}$ & \multirow{2}{*}{Ref.} \\ 
  & [GeV/nuc] & [cm$^{-2}$\,s$^{-1}$\,sr$^{-1}$\,(GeV/nuc)$^{-1}$] & \\
  \hline
 GAPS & $[0.05,0.25]$ & $2\times 10^{-6}$ &\cite{Aramaki:2015laa} \\  
 AMS-02 & 
 [0.2,  0.8] and [2.2, 4.2]
 & $4.5\times 10^{-7}$&\cite{Choutko2008} \\
 AMS-100 & $[0.1,8.0]$& $3\times 10^{-11}$ &\cite{Schael_2019}
\end{tabular}
\end{center}
    \caption{Expected antideuteron flux sensitivities of the three considered experiments.}
    \label{tab:sensitivities}
\end{table}

The marginalized flux allows one to compute the projected sensitivity for a parameter point $\vec x_\text{DM}$. 
To this end, we introduce the \emph{sensitivity factor} defined as the ratio of the expected marginal antideuteron flux to the respective sensitivity of the experiment:
\begin{equation}
    \label{eq:sensitivity_factor_dbar}
    \mathrm{Sensitivity\,\,factor} := \frac{\langle \Phi \rangle_{\overline{d}, E_\exper}}{\Phi_{\sensitivity, E_\exper}} \, ,
\end{equation}
where $\Phi_{\sensitivity, E_\exper}$ is the antideuteron flux sensitivity of the given experiment and $\langle \Phi \rangle_{\overline{d}, E_\exper}$ is the antideuteron flux at the experiment marginalized over all propagation parameters as described above. A sensitivity factor larger than one thus indicates that the experiment is sensitive to the DM signal. The sensitivity factor represents the expected number of events.

\section{Projections for experimental sensitivities}
\label{sec:results}

In this section, we use the tool-chain introduced in Sec.~\ref{sec:DRN} to derive the expected sensitivity factor for AMS-02, the upcoming GAPS experiment, and the future AMS-100 mission. We use the two propagation models DIFF.BRK and INJ.BRK introduced in Sec.~\ref{sec:prop} and consider two benchmark scenarios regarding the DM model. In the first scenario, annihilation is assumed to proceed exclusively into $b \bar b$; in the second, we consider a mixture of annihilation channels with relative contributions of a singlet scalar Higgs portal (SSHP) model with the Higgs portal coupling fixed by the requirement to explain the measured relic density, $\Omega h^2 =0.12$~\cite{Planck:2018vyg}, see \emph{e.g.}~\cite{Cuoco:2016jqt,DiMauro:2023tho}.\footnote{Note that the Higgs portal coupling only affects the relative contributions of the annihilation channels for DM masses above the $hh$ threshold, around $m_\text{DM}=m_h$. Below the threshold, annihilation always occurs via an $s$-channel Higgs such that the relative contributions depend on the DM mass and the couplings of the Higgs to SM particles only, \emph{i.e.}~they are independent of the DM type and the Higgs portal coupling.} 
Accordingly, both DM models have two free parameters only -- the DM mass, $m_\text{DM}$, and the total annihilation cross section, $\langle\sigma v\rangle_\text{tot}$. While varying the former, we consider two different choices for the latter. First, we consider the canonical annihilation cross section, $\langle\sigma v\rangle_\text{tot}= 3\times 10^{-26}\,\text{cm}^3/\text{s}$, required by thermal freeze-out. Second, we use the annihilation cross section that corresponds to the 95\% CL exclusion limit from antiprotons obtained in Ref.~\cite{Balan:2023lwg} for the respective propagation model considered, DIFF.BRK and INJ.BRK\@. Note that the exclusion limit does not reach below a DM mass of 10\,GeV, due to the large systematic uncertainties in the modeling of solar modulation and propagation at low rigidity~\cite{Kahlhoefer:2021sha}.

\begin{figure}
	\centering
	\begin{subfigure}[t]{0.49\textwidth}
        \centering
		\includegraphics[width=\textwidth]{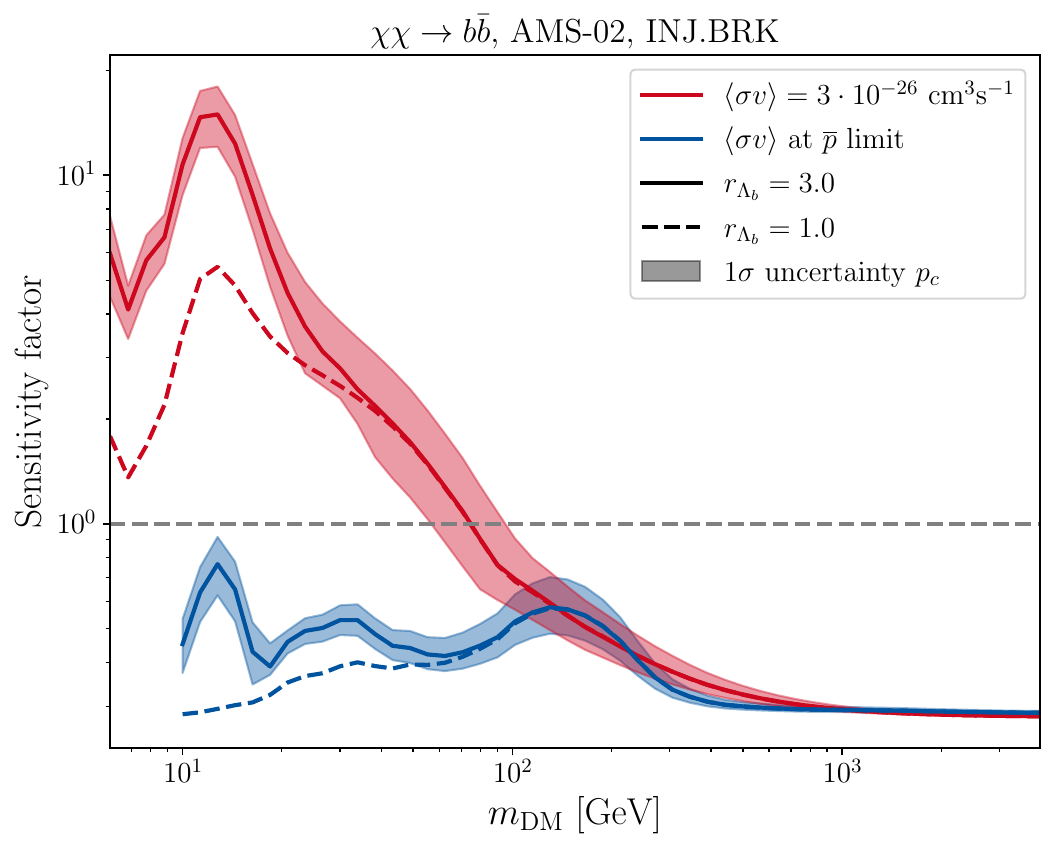}
        \label{}
	\end{subfigure}
	\hfill
     \begin{subfigure}[t]{0.49\textwidth}
        \centering
		\includegraphics[width=\textwidth]{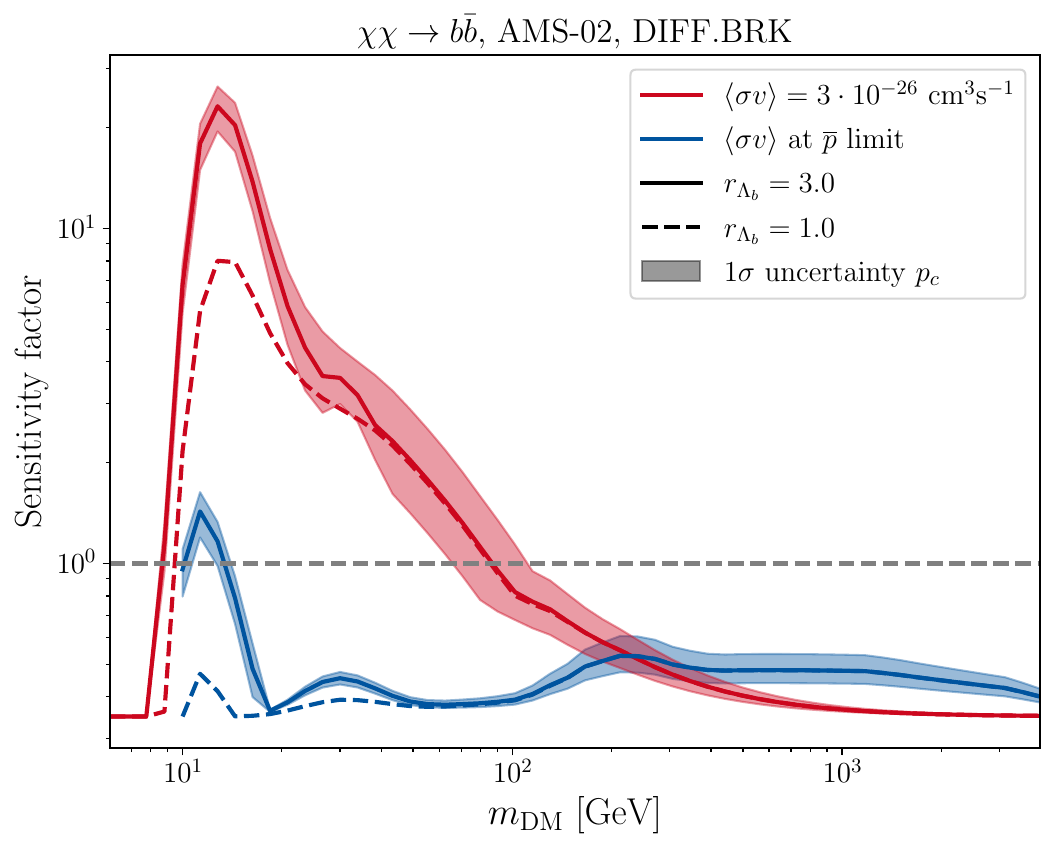}
        \label{}
	\end{subfigure}
    \begin{subfigure}[t]{0.49\textwidth}
        \centering
		\includegraphics[width=\textwidth]{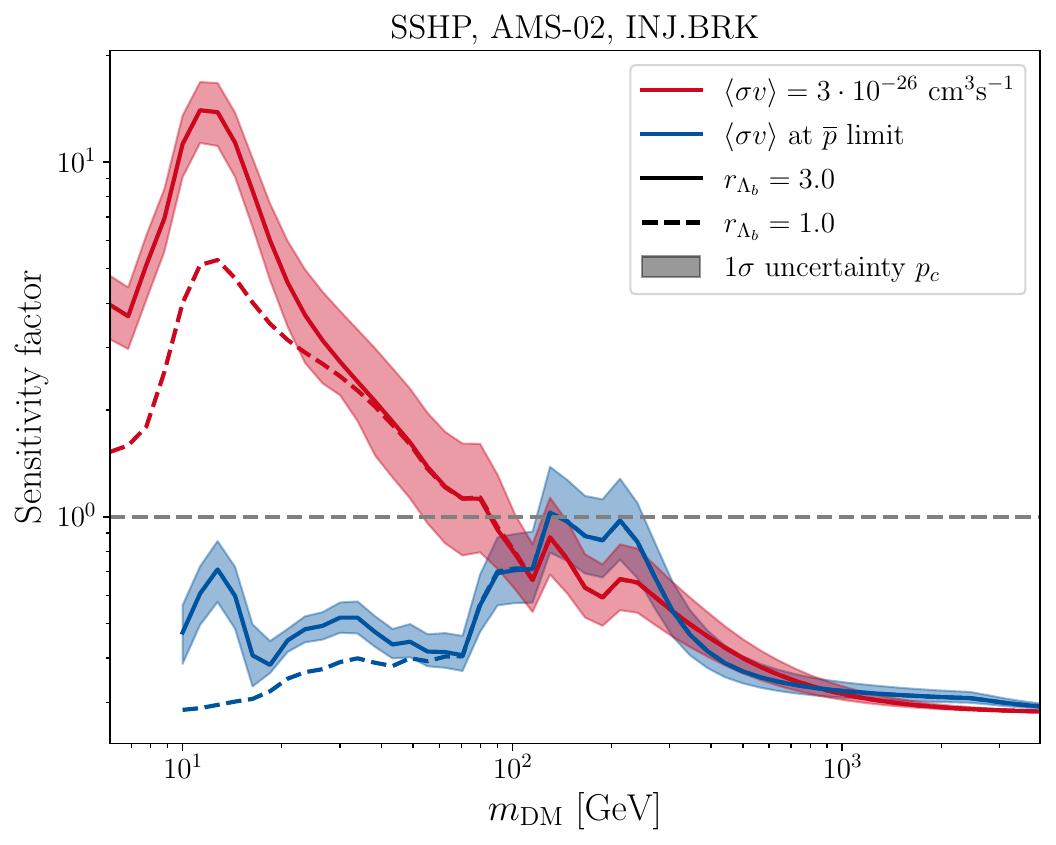}
        \label{}
	\end{subfigure}
	\hfill
     \begin{subfigure}[t]{0.49\textwidth}
        \centering
		\includegraphics[width=\textwidth]{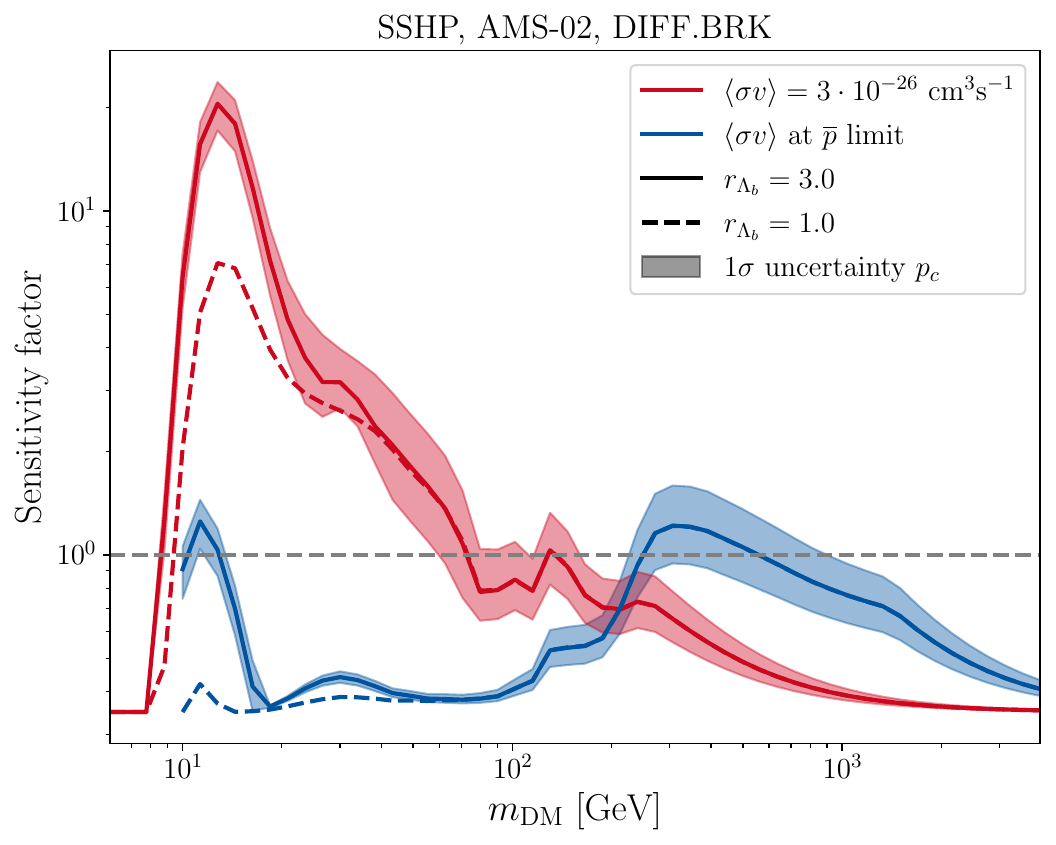}
        \label{}
	\end{subfigure}
	\caption{Sensitivity factor of AMS-02 experiment for annihilation into $b\bar b$ (upper panels) and the Higgs portal model (lower panels) within the propagation models INJ.BRK (left panels) and DIFF.BRK (right panels). The solid lines and shaded bands around them denote the results for a $\Lambda_b$ rescaling parameter of $r_{\Lambda_b}=3$ and the $1\sigma$ uncertainty from the determination of the coalescence momentum, respectively, while the dashed curves correspond to $r_{\Lambda_b}=1$. The red lines correspond to a fixed annihilation cross section while we use the annihilation cross section obtained from the antiproton exclusion limit for the blue lines.}
    \label{fig:sensAMS02}
\end{figure}

\begin{figure}
	\centering
	\begin{subfigure}[t]{0.49\textwidth}
        \centering
		\includegraphics[width=\textwidth]{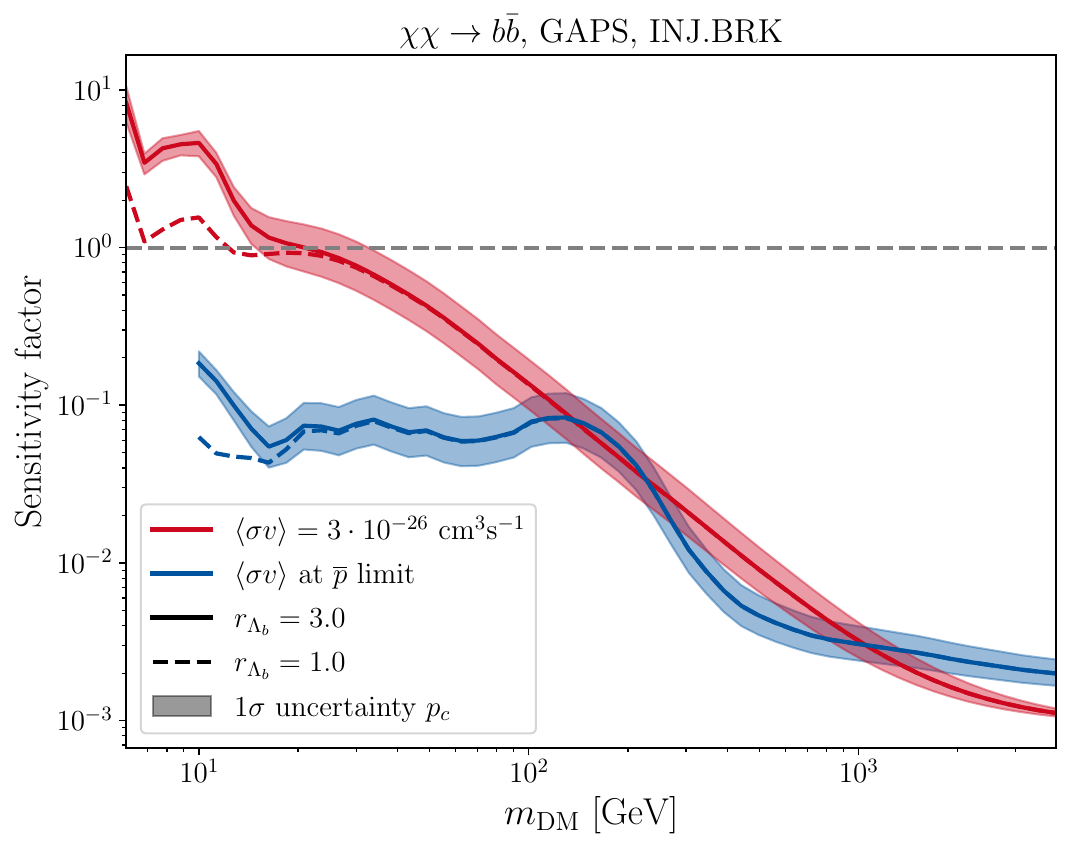}
        \label{}
	\end{subfigure}
	\hfill
     \begin{subfigure}[t]{0.49\textwidth}
        \centering
		\includegraphics[width=\textwidth]{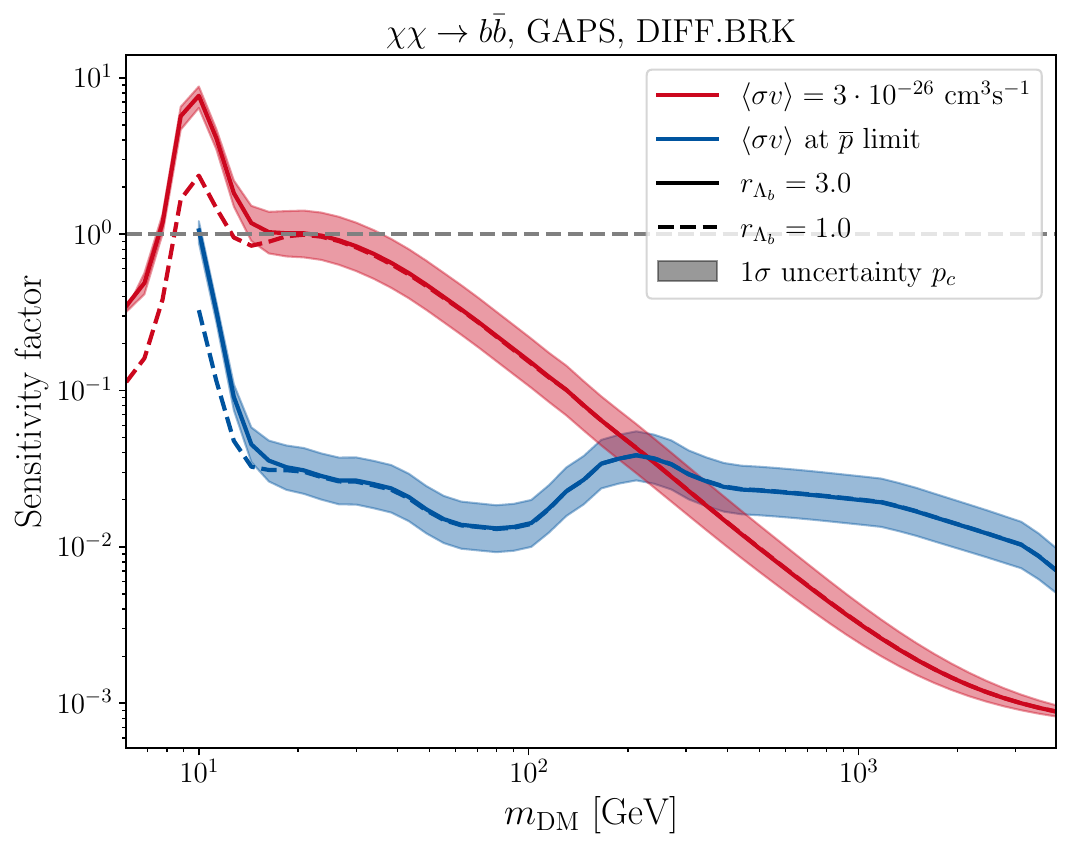}
        \label{}
	\end{subfigure}
	\begin{subfigure}[t]{0.49\textwidth}
        \centering
		\includegraphics[width=\textwidth]{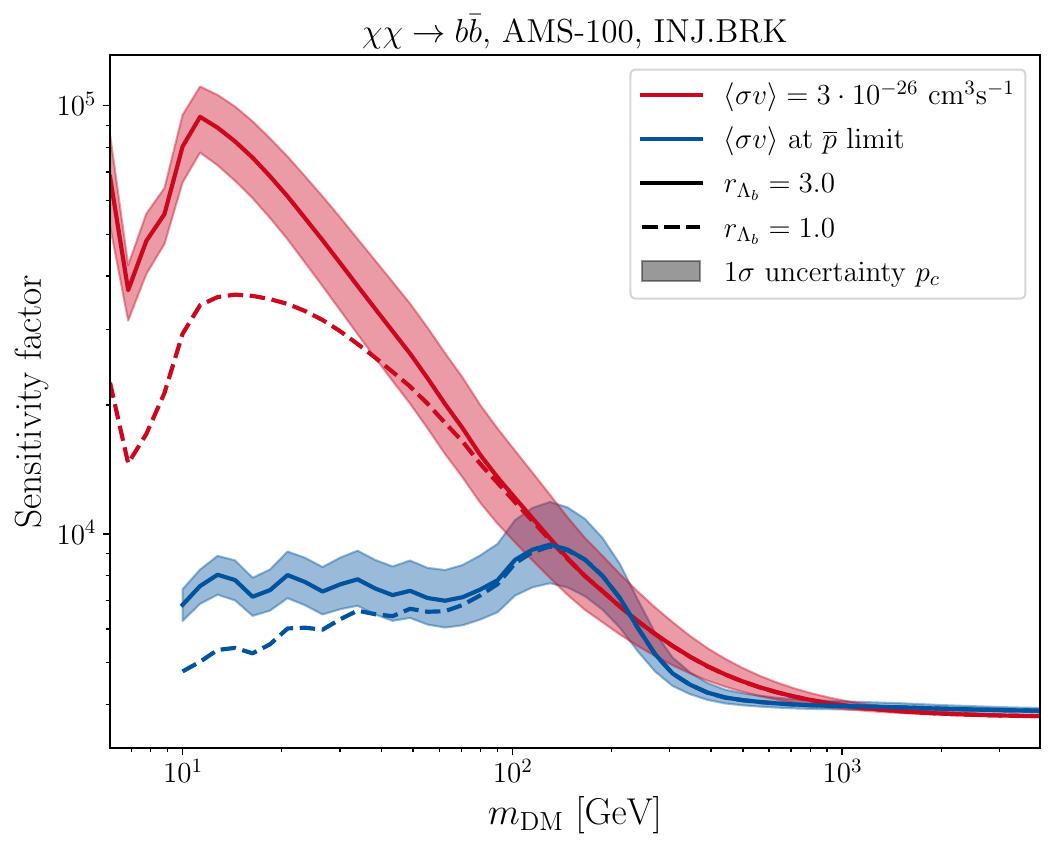}
        \label{}
	\end{subfigure}
	\hfill
     \begin{subfigure}[t]{0.49\textwidth}
        \centering
		\includegraphics[width=\textwidth]{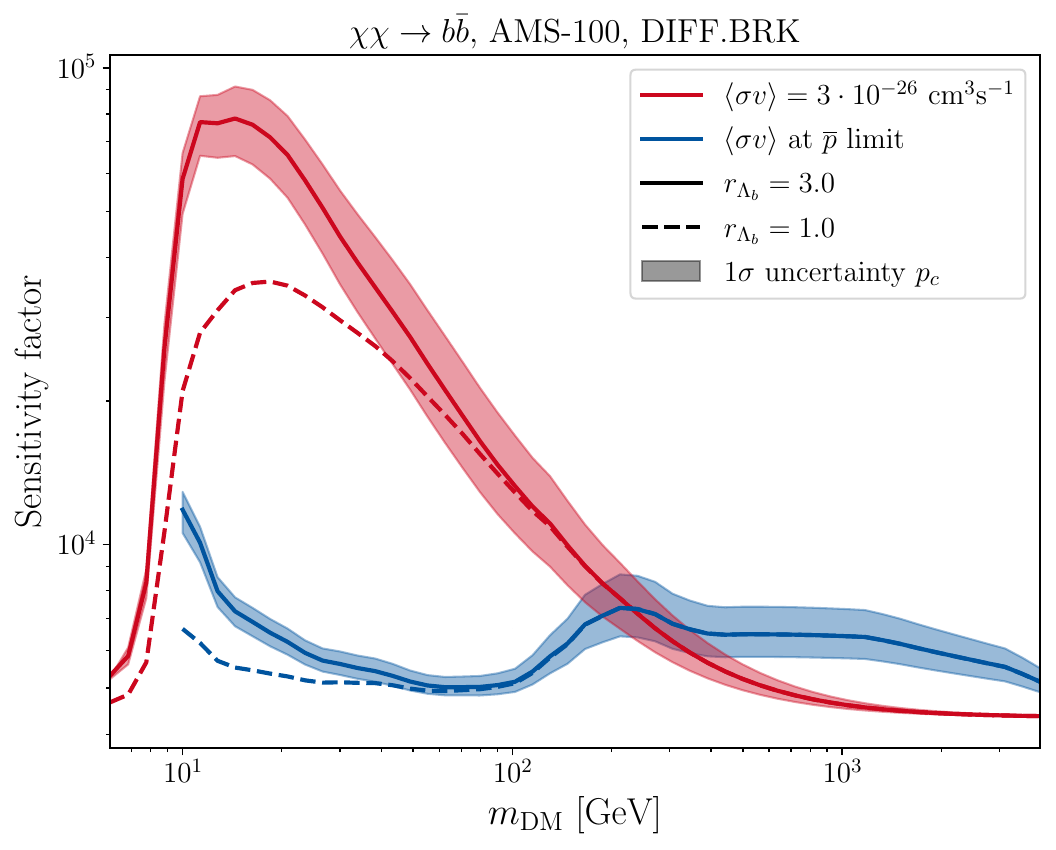}
        \label{}
	\end{subfigure}
	\caption{Sensitivity factor for GAPS (upper panels) and AMS-100 (lower panels) within the propagation models INJ.BRK (left panels) and DIFF.BRK (right panels). Here we consider annihilation into $b\bar b$ only. As in Fig.~\ref{fig:sensAMS02}, the solid lines and shaded bands denote the results for $r_{\Lambda_b}=3$ and the $1\sigma$ coalescence momentum uncertainty, respectively, while the dashed curves correspond to $r_{\Lambda_b}=1$.}
    \label{fig:sensGAPSAMS100}
\end{figure}

\begin{figure}
   \centering
    \includegraphics[scale=0.5]{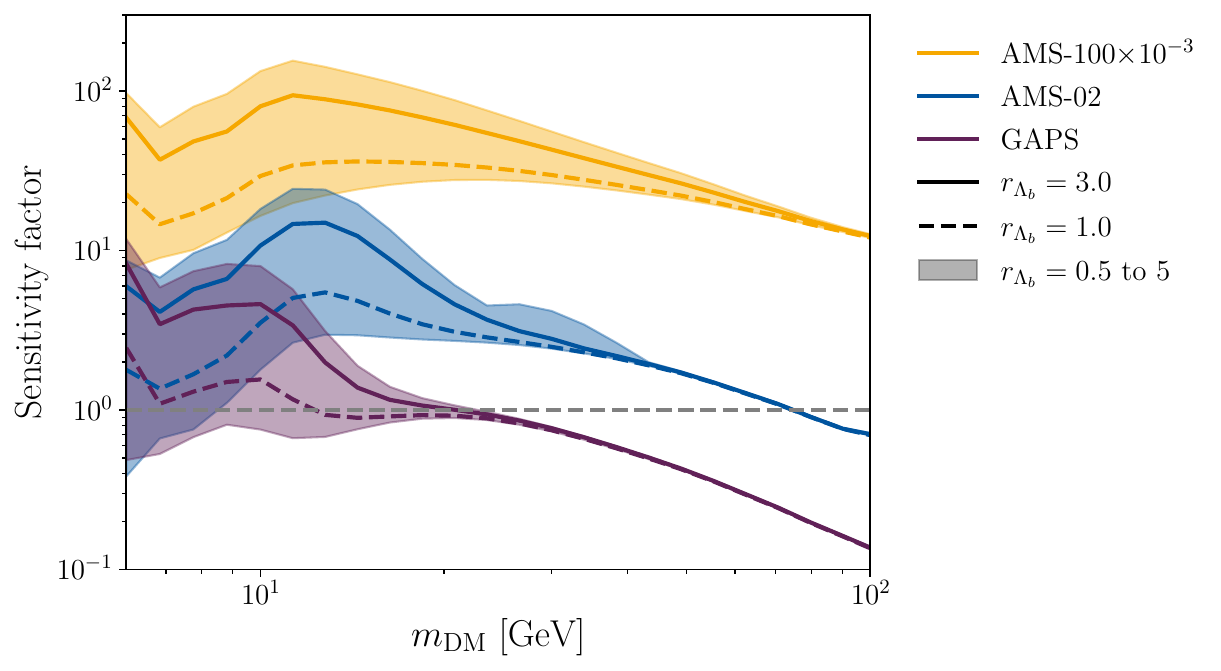}
    \caption{Sensitivity factor of all experiments in the INJ.BRK model for DM annihilation into $b \bar{b}$ with varying rescaling of the $\Lambda_b$ production rate $r_{\Lambda_b}$ assuming a thermal cross section of $\langle \sigma v \rangle = 3 \times 10^{-26}\, \mathrm{cm}^3 \mathrm{s}^{-1}$. Note that for better display the result for AMS-100 is scaled down by a factor of $10^{-3}$.}
    \label{fig:sensLambda}
\end{figure}

The resulting sensitivity factors are shown in Figs.~\ref{fig:sensAMS02} and \ref{fig:sensGAPSAMS100} for a DM mass between 5\,GeV and 4\,TeV\@. The red and blue curves correspond to the sensitivity factors assuming a thermal cross section and cross section at the 95\% CL exclusion limits from AMS-02 antiprotons, respectively. For masses above 10\,GeV where the antiprotons exclusion limit exists, the region above the blue line is excluded (within the assumptions of that analysis). In particular, these constraints exclude the thermal annihilation cross section for DM masses between 10\,GeV and at least around 100\,GeV.\footnote{Note that exclusion of the thermal annihilation cross section, $\langle \sigma v\rangle \simeq 3\times 10^{-26}\,\text{cm}^2/\text{s}$, does not necessarily imply an exclusion of a cosmologically viable DM model in the respective mass range. The annihilation cross section can be velocity-dependent, for instance, in the vicinity of kinematic thresholds or resonances, or a non-standard cosmological history can alter the required value of the thermal cross section. Therefore, the region above or below the red curve can still be considered phenomenologically relevant.} The upper edge of this mass range is strongly dependent on the propagation model. Within the DIFF.BRK model, it extend up to above 200\,GeV while in the INJ.BRK model it reaches up to around a TeV, however, except for a small window between roughly 100 and 200\,GeV, where the INJ.BRK model can accommodate the thermal cross section. The size of this window depends on the DM model and is larger for the considered Higgs portal model, where annihilation into $W^+W^-, ZZ$, and $hh$ are the most important annihilation channels in this mass range.

Considering both propagation and DM models,
AMS-02 shows sensitivity to the parameter region consistent with the antiproton exclusion in four distinct DM mass regions assuming our nominal value for $r_{\Lambda_b}$: first, and trivially, for $m_\text{DM}<10\,$GeV, where no antiproton exclusion limit exists; secondly, in a small mass region between 10 and 20\,GeV, however, only for the DIFF.BRK propagation model; thirdly, for the INJ.BRK propagation and Higgs portal DM model in the window 100--200\,GeV; and finally, for the DIFF.BRK propagation and Higgs portal DM model in the region above 200\,GeV\@. This results from the weakening antiproton limit above 200\,GeV in the INJ.BRK propagation model, which corresponds to testing larger values of $\langle \sigma v \rangle$. However, as the DM mass increases further, the antideuteron flux in the considered low-energy bins decrease more drastically, eventually leading to a loss of sensitivity for DM masses above roughly $ 600\,$GeV\@. 
Note that for dark matter masses below 50\,GeV, $b\bar b$ is the dominant annihilation channel in this model with a relative contribution of more than 80\%. Hence, in this region, the sensitivities between the two DM models are almost identical. Towards larger masses, the composition of annihilation channels deviates significantly and the singlet scalar Higgs portal model provides the higher sensitivity factor with $W^+W^-, ZZ$, and $hh$ being the dominant channels.

Our results indicate that a detection of around 7 antideuteron events at AMS-02~\cite{Ting:2023CERNSeminar} is consistent with limits from antiprotons for masses $m_\text{DM}<10\,$GeV only, assuming the expected sensitivity from Tab.~\ref{tab:sensitivities}. However, both up-to-date information on the AMS-02 sensitivity as well as a publication of the finding is still pending such that this observation is not conclusive.
Furthermore, we remind the reader that our choice of a relatively small  half-height of the diffusion volume, $z_h=4\,$kpc, provides a conservative estimate of the sensitivity factor introducing another source of uncertainty, see Sec.~\ref{sec:prop}.

The projected sensitivity factors of GAPS and AMS-100 are shown in the upper and lower panels of Fig.~\ref{fig:sensGAPSAMS100}, respectively, focusing on annihilation into $b\bar b$. The former will soon provide an important independent test of the existence of low-energy antideuterons. Its sensitivity can compete with AMS-02, in particular, in the aforementioned region $m_\text{DM}<10\,$GeV, due to its sensitivity to lower-energy antideuterons, see Fig.~\ref{fig:sensLambda} for a direct comparison. 
The projected sensitivity of the planned future mission AMS-100 is superior providing a sensitivity factor of more than three orders of magnitude above the one of AMS-02 and GAPS conclusively testing the thermal cross section up to a DM mass of around a TeV above which secondary antideuterons start to become relevant providing a background for the DM signal.\footnote{Note that AMS-02 and GAPS do not show sensitivity to secondary antideuterons. Accordingly, to good approximation, we can consider the search for antideuterons from DM background-free. 
In contrast, AMS-100 does provide sensitivity to secondary antideuterons. In the region of very low antideuteron yields from DM, the sensitivity factor shown reflects the sensitivity to detect any antideuteron while the sensitivity to detect a DM contribution in this region would require a different analysis, which is, however, beyond the scope of this work.}

Our analysis also sheds light on the sensitivity factor's dependence on the coalescence  parameters. The shaded bands in Figs.~\ref{fig:sensAMS02} and \ref{fig:sensGAPSAMS100} denote the $1\sigma$ uncertainty in the coalescence momentum $p_\text{c}$. The bands shrink in the region of small sensitivity factors (typically at large DM masses) as well as towards small DM masses. The former effect is simply due to a non-negligible -- and eventually dominating -- contribution from secondary emission for which we do not consider a similar variation in their production rate. The latter occurs due to the weaker $p_\text{c}$-dependence of anitdeuteron production from $\Lambda_b$ baryons which constitutes a large contribution in this rage. This can be seen from the 
dependence of the sensitivity factor on the $\Lambda_b$ production rate indicated in the plots: the solid lines in Figs.~\ref{fig:sensAMS02} and \ref{fig:sensGAPSAMS100} denote the nominal value $r_{\Lambda_b}=3$ while the dashed lines show the result without rescaling the $\Lambda_b$ production rate, $r_{\Lambda_b}=1$. As antideuteron production through $\Lambda_b$ baryons only affects a small part of the annihilation spectrum for $b\bar b$ towards large $x$ (see upper left panel of Fig.~\ref{fig:SourceSpectra}), the rescaling of $r_{\Lambda_b}$ only affects small DM masses for which the `$\Lambda_b$-shoulder' overlaps with the energy bin of the search. Additionally, we highlight the $r_{\Lambda_b}$-dependence in Fig.~\ref{fig:sensLambda}, where we focus on INJ.BRK model for DM annihilation into $b\bar b$. The shaded bands show the variation of $r_{\Lambda_b}$ between 0.5 and 5, indicating the high importance of a precise measurement in the region of small DM masses, below 20\,GeV for GAPS and progressively higher for AMS-02 and AMS-100 due to their wider energy ranges.

\section{Conclusions}
\label{sec:concl}

Low-energy cosmic-ray antideuterons are a smoking-gun signature of DM annihilation in our Galaxy. In this work, we provided a careful analysis of this signature in light of the currently running AMS-02 experiment, the upcoming GAPS experiment, and the future AMS-100 mission.
Revisiting the production of antideuterons from DM annihilation through a Monte Carlo simulation, we included uncertainties of the coalescence model, specifically the coalescence momentum inferred from LEP data, as well as the $\Lambda_b$ baryon production rate affecting the `$\Lambda_b$-shoulder' for annihilation into $b\bar b$. In addition, and importantly, we took into account uncertainties from cosmic-ray propagation affecting the cosmic-ray antideuteron yield measured near Earth by varying the propagation parameters within two distinct propagation models.

To enable the inclusion of these uncertainties via marginalization (or profiling) over the corresponding parameters in the limit-setting procedure in future interpretations of antideuteron fluxes, we developed \DRN, a fast neural-network based emulation of the antideuteron production and propagation process suitable for the large number of evaluations in such an analysis. Our tool provides the propagated antideuteron spectra directly from the (admixture of) annihilation channels and the cross section in a fast and accurate manner.
We have employed a recurrent neural network suitable for describing binned spectra with strong correlations among neighboring energy bins. 
The network performs sufficiently well, with 68\% of transformed fluxes having a relative difference of at most $6 \times 10^{-4}$ in most bins, resulting in a relative error of $\mathcal{O}(10^{-2})$ in the actual flux.

The fast emulation of propagated antideuteron fluxes described here enables the study of a wide range of DM models which would otherwise demand excessive computational resources due to two key factors. First, Monte Carlo-based computations of sufficiently smooth injection spectra require simulating $\mathcal{O}(10^8)$ events for each DM model parameter point. Second, reliably predicting propagation effects involves solving the diffusion equation with tools like \textsc{Galprop}. This is a computationally intensive task, especially when accounting for propagation uncertainties in the limit-setting, which requires numerous evaluations to marginalize over the high-dimensional propagation parameter space. \DRN\ computes the marginalized flux in mere seconds per CPU, in stark contrast to the significantly longer computation times required by the traditional methods mentioned above. As a result, the tool is well-suited for performing extensive DM parameter scans.

We demonstrated our tool by computing the expected sensitivity factor for AMS-02, GAPS, and AMS-100 for the two different state-of-the-art propagation models and for two different DM models -- pure annihilation into $b \bar b$ and a Higgs portal model.
Our analysis reveals that AMS-02 provides sensitivity to annihilation into $b\bar b$ that is consistent with limits from cosmic-ray antiprotons only for small DM masses, $\mDM\lesssim20\,$GeV, whereas in the singlet scalar Higgs portal model, a further mass window around (above) 200\,GeV shows sensitivity while being consistent with those constraints assuming the INJ.BRK (DIFF.BRK) propagation model.
If AMS-02 confirms the measurements of ${\cal O}(7)$ low-energy antideuterons, it will only be consistent with antiproton limits for $m_\text{DM}<10\,$GeV when assuming the expected AMS-02 sensitivity within our propagation models. Interestingly, in this range, GAPS can compete with AMS-02 providing a similar sensitivity to the models considered, and thus allowing a critical test of this scenario. While the sensitivity of antideuteron searches with both experiments does not significantly exceed the one of current antiproton searches, they still provide a crucial independent probe, given their significantly smaller backgrounds. Ultimately, the ambitious AMS-100 mission will offer superior sensitivity testing the entire parameter range up to the TeV scale where, however, the secondary antideuteron background becomes significant.

\section*{Acknowledgments}

We thank Sowmiya Balan, Henning Gast, and Felix Kahlhoefer
for discussions. J.H.\ acknowledges support from the Alexander von Humboldt Foundation through the Feodor Lynen Research Fellowship for Experienced Researchers and the Feodor Lynen Return Fellowship. LR's research is supported by the DFG under grant 396021762 - TRR 257: Particle Physics Phenomenology after the Higgs discovery. M.Ko.~is supported by the Swedish Research Council under contracts 2019-05135 and 2022-04283, and by the European Research Council under grant 742104. This work was carried out in part at the Aspen Center for Physics, supported by National Science Foundation grant PHY-2210452.
Simulations were performed using computing resources provided by RWTH Aachen University under project rwth0754. 
This project used computing resources from the Swedish National Infrastructure for Computing (SNIC) and National Academic Infrastructure for Supercomputing in Sweden (NAISS) under project Nos.~2021/3-42, 2021/6-326, 2021-1-24 and 2022/3-27, partially funded by the Swedish Research Council through grant no.~2018-05973.

\begin{appendix}
\section{Spatial separation}\label{app:spatialsep}

We define the spatial separation $\delta r$ between $\bar p$ and $\bar n$ as their minimum distance along their trajectories in the $\bar d$ rest frame~\cite{Fornengo:2013osa}. 

To boost a four-vector from the laboratory frame (denoted by a prime) to the $\bar d$ rest frame (vectors not primed), we apply the Lorentz transformation
\begin{equation}
\Lambda = 
	\begin{pmatrix}
		\gamma & -\gamma\beta_i \\
		-\gamma\beta_j & \delta_{ij} + (\gamma-1) \frac{\beta_i\beta_j}{\beta^2} 
	\end{pmatrix}\,,
\end{equation}
where $i,j$ denote the spatial components. For the $\bar d$ four-momentum, $p_{\bar d}'= p_{\bar p}'+p_{\bar n}'$,
by definition
\begin{equation}
\Lambda p_{\bar d}' = \begin{pmatrix} (p_{\bar d}')^2 \\ \vec{0} \end{pmatrix}\,,
\end{equation}
and hence
\begin{equation}
\gamma = \frac{p_{\bar d}'^{\,0}}{(p_{\bar d}')^2}\,,\quad \beta_i = \frac{p_{\bar d}'^{\,i}}{p_{\bar d}'^{\,0}}\,.
\end{equation}
The velocity of $\bar p$ in the $\bar d$ rest frame is
\begin{equation}
v_{\bar p}^{\,i}= \frac{p_{\bar p}^{\,i}}{p_{\bar p}^{\,0}}\,,
\end{equation}
and analogous for the $\bar n$. 
Given the four-vector 
for the $\bar p$ production  
in the $\bar d$ rest frame, $x_{\bar p,0}$,
we can parameterize its trajectory as
\begin{equation}
\vec{x}_{\bar p}(t) = \vec{x}_{\bar p,0} + \vec{v}_p (t - x_{\bar p,0}^0)\,,
\end{equation}
and analogous for the $\bar n$ such that minimal distance along their trajectories is
\begin{equation}
\Delta r^2 = \big| \vec{x}_{\bar p}(t_{\min}) - \vec{x}_{\bar n}(t_{\min}) \big|^2 = \vec{D}^2 + 2t_{\min} \vec{D}\!\cdot \!\vec{v}  + t_{\min} \vec{v}^2 \,,
\end{equation}
with
\begin{equation}
\vec{D} = \vec{x}_{\bar p,0} - \vec{x}_{\bar n,0} + \vec{v}_{\bar p} x_{\bar p,0}^0 - \vec{v}_{\bar n} x_{\bar n,0}^0\,,\quad \vec v = \vec{v}_{\bar p} - \vec{v}_{\bar n}\,,
\end{equation}
and
\begin{equation}
t_{\min} = \max\left( - \frac{\vec{D}\!\cdot \!\vec{v} }{ \vec{v}^2},\, x_{\bar p,0}^0, \,x_{\bar n,0}^0 \right)\,.
\end{equation}

\section{Calibration of  coalescence model}\label{app:LEPfit}

The coalescence momentum $p_c$ is a free parameter of the coalescence model that must be determined by fitting the simulated antideuteron production rate to experimental data. Antideuteron production has been measured in various laboratory experiments, for example by hadronic processes such as $p p \rightarrow \bar{d}$ at the LHC or leptonic processes such as $e^+ e^- \rightarrow \bar{d}$ at LEP \cite{Fornengo:2013osa}. To mimic the conditions of DM annihilation as closely as possible (i.e., ~non-hadronic initial states), we only use data from $e^+ e^- \rightarrow \bar{d}$. Accordingly, we use the production rate of antideuterons in $e^+ e^-$ collisions at the $Z$ resonance per hadronic $Z$ decay measured by ALEPH, $R_{\bar{d}} = \big(5. 9 \pm 0.5 \, (\mathrm{sys.})\big) \times 10^{-6}$ in the momentum range $0.62 < p < 1.03\,$GeV and a polar angle in the interval $|\cos{\theta}| < 0.95$~\cite{ALEPH:2006qoi}. The systematic error is due to the uncertainty in the antideuteron detection efficiency. 

The total number of measured antideuterons from primary interactions is 11~\cite{ALEPH:2006qoi}.
We therefore assume Poisson statistics to compute the upper and lower bounds of the corresponding 68\% and 95\% confidence intervals~\cite{Workman:2022ynf}. These limits are then divided by the observed number of hadronic $Z$ decays and the antideuteron detection efficiency to determine the total uncertainty in the antideuteron production rate. We obtain $R_{\bar{d}} = \big(5.9 \, ^{+2.5}_{-1.8} \, (\mathrm{stat.}) \pm 0.5 \, (\mathrm{sys.}) \big)\times 10^{-6}$ at 68\% confidence level.

\begin{figure}
    \centering
    \includegraphics[scale=0.5]{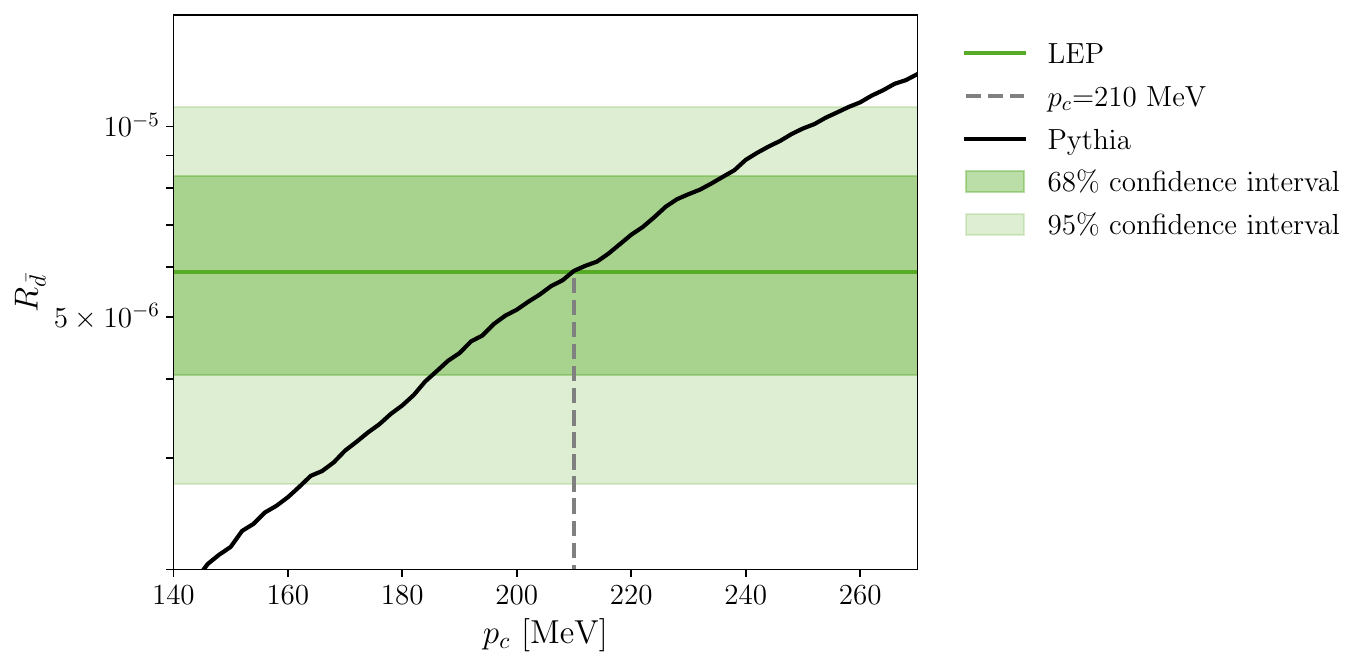}
    \caption{Rate of antideuterons measured by LEP and simulated with Pythia as a function of coalescence momentum to determine the upper and lower bounds on the coalescence momentum.}
    \label{fig:LEP_fit}
\end{figure}

To simulate the hard process of $e^+ e^-$ collisions at LEP, we use the MC event generator \textsc{MadGraph\_aMC@NLO}~\cite{Alwall:2014hca}. Specifically, we generate $10^8$  events for the process $e^+ e^-\to Z\to q \bar q$ at the Z resonance. For the simulation of showering and hadronization as well as the analysis of events, we use the same tool-chain as for DM annihilation.
We evaluate the resulting simulated antideuteron production rate as a function of $p_c$ and compare it to the above-mentioned LEP measurement. The results are presented in Fig.~\ref{fig:LEP_fit} and lead to the best-fit coalescence momentum and 68\% (95\%) confidence level intervals:
\begin{equation}
    p_c = 
    210 \,{}^{+27}_{-25}\,
    \,\big({}^{+48}_{-47}\big) \,
    \, \mathrm{MeV}\,.
    \label{eq:p_c_value}
\end{equation}
This result is similar to the one found in Ref.~\cite{Winkler:2020ltd}, $p_c = 215^{+19}_{-23}$\,MeV.

\section{Computation of marginalized flux}\label{app:LikeSen}

We compute the expected marginalized flux by using the source term from the neural network and the \textsc{Galprop} code to propagate the antideuterons to Earth. We treat the marginalization as follows: 
\begin{align}
    \langle \Phi \rangle_{\dbar, E_\exper} (\xDM) &= \int \diff \tprop \, \,p_\mathrm{DM}(\tprop, \xDM) \, \Phi_{\dbar, E_\exper} (\tprop, \xDM)\\
    & = \int \diff \tprop \, \,p_\mathrm{DM}(\tprop, \xDM) \, \frac{p(\tprop)}{p(\tprop)} \, \Phi_{\dbar, E_\exper} (\tprop, \xDM)\\
    & = \underbrace{\int \diff \tprop \, \,p(\tprop)}_{\sim \sum_i} \, \underbrace{\frac{p_\mathrm{DM}(\tprop, \xDM)}{p(\tprop)}}_{(\ast)} \, \Phi_{\dbar, E_\exper} (\tprop, \xDM)\\
    &\approx \underbrace{\frac{Z}{Z_\mathrm{DM}}}_{(\ast \ast)} \, \sum_i \Phi_{\dbar, E_\exper} (\tpropi, \xDM) \, \frac{\mathcal{L}_\mathrm{DM} (\tpropi, \xDM)}{\mathcal{L} (\tpropi)}\\
    & =  \left(\sum_i \frac{\mathcal{L}_\mathrm{DM} (\tpropi, \xDM)}{\mathcal{L} (\tpropi)}\right)^{-1} \, \sum_i \Phi_{\dbar, E_\exper} (\tpropi, \xDM) \, \frac{\mathcal{L}_\mathrm{DM} (\tpropi, \xDM)}{\mathcal{L} (\tpropi)} \, .
\end{align}
Here $p$ denotes the posterior, $\pi$ the prior, and $Z$ the evidence, i.e., the normalization of the posterior.
We start with the definition of the marginal flux and rewrite it so that we can approximate the integral as a sum over the posterior sample of the propagation parameters. 

In the third step, we can also rewrite the posterior ratio using Bayes' theorem as follows:
\begin{equation}
    (\ast) = \frac{p_\mathrm{DM}(\tprop, \xDM)}{p(\tprop)} = \frac{Z}{Z_\mathrm{DM}} \, \frac{\pi(\tprop)}{\pi(\tprop)} \, \frac{\mathcal{L}_\mathrm{DM} (\tprop, \xDM)}{\mathcal{L}_\mathrm{DM} (\tprop)} \, .
\end{equation}
Finally, we can express the evidence ratio as 
\begin{align}
    (\ast \ast) & = \frac{Z_\mathrm{DM}}{Z}\\
    & = \frac{\int \diff \tprop \pi (\tprop) \mathcal{L}_\mathrm{DM}(\tprop, \xDM)}{Z} \\
    & = \int \diff \tprop \frac{\pi (\tprop) \mathcal{L}  (\tprop)}{Z} \frac{\mathcal{L}_\mathrm{DM}(\tprop, \xDM)}{\mathcal{L}  (\tprop)} \\
    & = \int \diff \tprop p(\tprop) \frac{\mathcal{L}_\mathrm{DM}(\tprop, \xDM)}{\mathcal{L}  (\tprop)} \\
    &\approx \sum_i \frac{\mathcal{L}_\mathrm{DM} (\tpropi, \xDM)}{\mathcal{L} (\tpropi)} \, .
\end{align}
This allows us to express the marginal flux as a weighted sum over the posterior sample of fluxes at Earth. The weights are given by the likelihood ratio of the DM model compared to the mere background (\emph{i.e.}~secondary emission). For this, we can use the AMS-02 antiproton likelihood ratio, computed with the tool \textsf{pbarlike}, published together with~\cite{Balan:2023lwg}.

\end{appendix}

\bibliographystyle{JHEP_improved}
\bibliography{bibliography}

\end{document}